\documentclass{aa}
\usepackage[varg]{txfonts}
\usepackage{graphicx}
\usepackage[version=3]{mhchem} 
\usepackage{longtable}
\bibpunct{(}{)}{;}{a}{}{,} 
\usepackage{color}

\begin{document}
\title{The role of environment and gas temperature in the formation of multiple protostellar systems: molecular tracers}

\author{N. M. Murillo\inst{1} \and E. F. van Dishoeck\inst{1,2} \and J. J. Tobin\inst{3} \and J. C. Mottram\inst{4} \and A. Karska\inst{5}}

\institute{Leiden Observatory, Leiden University, P.O. Box 9513, 2300 RA, Leiden, the Netherlands \\ \email{nmurillo@strw.leidenuniv.nl}
        \and Max-Planck-Institut f\"{u}r extraterrestrische Physik, Giessenbachstra\ss e 1, 85748, Garching bei M\"{u}nchen, Germany
        \and Homer L. Dodge Department of Physics and Astronomy, University of Oklahoma, 440 W. Brooks Street, Norman, Oklahoma 73019, USA
    	\and Max Planck Institute for Astronomy, K\"{o}nigstuhl 17, 69117, Heidelberg, Germany
        \and Centre for Astronomy, Nicolaus Copernicus University, Faculty of Physics, Astronomy and Informatics, Grudziadzka 5, 87100, Torun, Poland}

\abstract
{Simulations suggest that gas heating due to radiative feedback is a key factor in whether or not multiple protostellar systems will form. Chemistry is a good tracer of the physical structure of a protostellar system, since it depends on the temperature structure.}
{To study the relationship between envelope gas temperature and protostellar multiplicity.}
{Single dish observations of various molecules that trace the cold, warm and UV-irradiated gas are used to probe the temperature structure of multiple and single protostellar systems on 7000 AU scales.}
{Single, close binary and wide multiples present similar current envelope gas temperatures, as estimated from \ce{H2CO} and \ce{DCO+} line ratios. The temperature of the outflow cavity, traced by \ce{c-C3H2}, on the other hand, shows a relation with bolometric luminosity and an anti-correlation with envelope mass. Although the envelope gas temperatures are similar for all objects surveyed, wide multiples tend to exhibit a more massive reservoir of cold gas compared to close binary and single protostars.}
{Although the sample of protostellar systems is small, the results suggest that gas temperature may not have a strong impact on fragmentation. We propose that mass, and density, may instead be key factors in fragmentation.}


\titlerunning{Temperature and multiplicity}
\authorrunning{Murillo et al.}
   
\maketitle

\section{Introduction}
\label{sec:intro}

Multiple protostellar systems are widely thought to be formed through fragmentation of the cloud core and/or disk within which they form.
This process is expected to either be induced by turbulence (e.g., \citealt{offner2010}) or through instabilities in the disk that can lead to fragmentation of the disk material (e.g., \citealt{stamatellos2009a,kratter2010}).
Each mechanism is proposed to produce multiple protostellar systems, but since they operate on different spatial scales (disks: $\sim$100 AU, vs. cloud core: $\sim$1000 AU) they result in different separations between the sources. 
Turbulent fragmentation predicts the initial formation of wide companions, whereas disk fragmentation can produce close companions, on scales of the disk radius.
The time when these processes occur, upon initial collapse or after one source has formed, can also alter the resulting multiple protostellar system and its evolution.
Observational studies of massive star formation provide conflicting evidence regarding the role of turbulence in core fragmentation (e.g., \citealt{wang2014,palau2015,beuther2018}).
Another potentially relevant factor in regulating fragmentation may be magnetic support. This mechanism suggests that magnetic fields reduce the number of fragments formed in a cloud core \citep{commercon2010,hennebelle2011}. Observations of fragmentation in massive dense cloud cores \citep{tan2013,fontani2018}, where both high- and low-mass protostars can form, suggest that magnetic fields can shape the fragment mass and distribution. The role that magnetic support plays in low-mass multiple star formation is not yet entirely understood.

The factors that enhance fragmentation of cloud cores need to be studied in order to understand how multiple protostellar systems form.
Radiative feedback and gas heating have been raised as key factors in the fragmentation of protostellar cores (e.g., \citealt{krumholz2006,bate2012,krumholz2014}).
Simulations and models show that fragmentation is suppressed by heated gas due to the increase in the thermal Jeans mass needed for collapse.
An accreting protostar heats up its surrounding gas, even as early as the first collapse of the core \citep{boss2000,whitehouse2006}. Thus it is expected that fragmentation can be considerably suppressed even as the protostellar object forms.
Numerical simulations show that as stars begin to form they can heat surrounding gas out to a few thousand AU, with the gas being continuously heated out to larger expanses as more objects form \citep{bate2012}. 
This is expected to considerably reduce fragmentation, and consequently the formation of multiple protostellar systems on envelope scales (few thousand AU).
Models considering the effect of accretion luminosity on the temperature structure of cloud cores suggest that cores can be heated above 100 K out to a few hundred AU, and 30 K out to a few thousand AU \citep{krumholz2014}. This is thought to significantly hinder cloud core fragmentation and the consequent formation of multiple stellar systems \citep{krumholz2006}.

Observations of young low-mass embedded protostellar systems, however, seem to show a different picture.
Construction of the SEDs of all known embedded protostellar systems in the Perseus star forming region (d $\sim$ 235 pc, \citealt{hirota2011}) found, for separations larger than 7$\arcsec$, that higher order multiples have a tendency for one of the sources to be at a different evolutionary stage than the rest, i.e. non-coeval systems \citep{murillo2016}.
For these non-coeval systems to occur, one of the sources was most likely formed after the other protostars were formed.
During the star formation process, episodic accretion bursts produce quiescent phases, lasting on the order of 10$^{3-4}$ yr \citep{scholz2013,visser2015}, which allow enough time for the envelope to cool and thus be more conducive to fragmentation and collapse.
However, objects undergoing episodic accretion do not always show multiplicity (e.g., very low luminosity objects, VeLLOs; \citealt{hsieh2018}), and not all multiples present signatures of accretion bursts (e.g., \citealt{frimann2017}).
Furthermore, the recently fragmented circumbinary disk of the deeply embedded protostellar system L1448 N \citep{tobin2016N} suggests that instabilities could overcome heating-suppressed fragmentation, since this disk is most certainly heated by both the central binary and through accretion.

Observational evidence for gas and dust heating on scales of a few thousand AU by UV radiation escaping through outflow cavities comes from several lines of evidence. 
Multi-wavelength observations of dust emission around low-mass protostars suggest indeed elevated temperatures out to such distances (e.g., \citealt{hatchell2013,sicilia2013}). 
More relevant are measurements of the gas temperature, since at low cloud densities (10$^{4}$ cm$^{-3}$) gas and dust temperatures may be decoupled \citep{evans2001,galli2002}, and it is the gas temperature that enters the formulation for suppressing fragmentation \citep{offner2010}. 
The best diagnostics are the \ce{^{13}CO} mid-J lines, especially \ce{^{13}CO} J = 6--5, first demonstrated by \citet{spaans1995} and modeled in detail by \citet{visser2012}.
The use of \ce{^{13}CO} 6--5 as a temperature probe on extended scales has been quantified observationally by \citet{vankempen2009} and \citet{yildiz2013,yildiz2015}, showing temperatures of 30-50 K on scales of a few thousand AU.

Relating the temperature structure and multiplicity of a protostellar system can provide constraints on the temperature-fragmentation relation.
Simulations including radiative feedback (e.g., \citealt{krumholz2006,bate2012}) would lead us to expect that non-coeval multiple protostellar systems may form in much colder cloud cores than single and coeval binary protostellar systems, in order to have further fragmentation.
Thus, the temperature structure of protostellar systems needs to be characterized in order to test these models.
Because heating is time-dependent and we cannot observe the temporal history of protostellar envelope heating, protostellar objects at different evolutionary stages and having recently undergone processes such as accretion bursts and fragmentation need to be studied and compared.

Molecular excitation and chemistry provide an excellent tool to probe the temperature structure of a protostellar system.
Using selected molecules that trace the cold and warm envelope gas (e.g., \citealt{murillo2015,murillo2018}), it can be established how the gas heating is being distributed throughout the cloud core.
In addition, we use new \ce{^{13}CO} 6--5 data combined with existing \ce{^{13}CO} 3--2 spectra \citep{mottram2017} and \ce{^{13}CO} 10--9 spectra from the WILL survey \citep{mottram2017}, observed with the James Clerk Maxwell Telescope (JCMT) and \textit{Herschel Space Observatory} \citep{pilbratt2010}, respectively.
Relating this to the multiplicity and coevality\footnote{Coevality is here used as defined in \citet{murillo2016}, which is the relative evolutionary classes of sources in a multiple protostellar system.} can then provide information on how temperature affects fragmentation.

This work presents single-dish observations of embedded multiple and single protostellar systems aiming to address the relation between temperature and fragmentation at the envelope scale ($\sim$7000 AU).
The sample selection criteria for the protostellar systems studied in this work are described in Section~\ref{sec:sample}.
Section~\ref{sec:obs} describes the Atacama Pathfinder EXperiment (APEX; \citealt{gusten2006}) observations.
Results and analysis of the data obtained from the observations are given in Sections~\ref{sec:results} and \ref{sec:analysis}, respectively.
Comparison between single and multiple protostellar systems is made in Section~\ref{sec:diss}, considering evolutionary stage and whether they are located in a crowded or isolated environment.
The conclusions of this work are given in Section~\ref{sec:conc}, along with the resulting insight on the temperature-fragmentation relation.

\begin{table*}
	\centering
	\caption{Sample of protostellar systems}
	\begin{tabular}{c c c c c c c c}
		\hline \hline
		System & Source & RA & Dec & Separation ($\arcsec$) & Class & Region\tablefootmark{b} & $L_{\rm bol}$ (L$_{\odot}$) \\
		\hline
		\multicolumn{8}{c}{Wide multiples}\\
		\hline
		L1448N & A & 03:25:36.53 & +30:45:21.35 & ... & I & non-clustered & 5.88 $\pm$ 0.93 \\
		& B & 03:25:36.34 & +30:45:14.94 & 7.3 & 0 & & 2.15 $\pm$ 0.33 \\
		& C & 03:25:35.53 & +30:45:34.20 & 16.3 & 0 & & 1.22 $\pm$ 0.19 \\
		NGC1333 SVS13 & A & 03:29:03.75 & +31:16:03.76 & ... & I & clustered & 119.28 $\pm$ 18.31 \\
		& B & 03:29:03.07 & +31:15:52.02 & 14.9 & 0 & & 10.26 $\pm$ 1.57 \\ 
		& C & 03:29:01.96 & +31:15:38.26 & 34.7 & 0 & & 2.22 $\pm$ 0.34 \\
		NGC1333 IRAS5 & Per63 & 03:28:43.28 & +31:17:32.9 & ... & I & clustered & 1.38 $\pm$ 0.21 \\
		& Per52 & 03:28:39.72 & +31:17:31.9 & 45.7 & I & & 0.12 $\pm$ 0.02 \\
		NGC1333 IRAS7 & Per18 & 03:29:11.26 & +31:18:31.08 & ... & 0 & clustered & 4.77 $\pm$ 0.73 \\
		& Per21 & 03:29:10.67 & +31:18:20.18 & 13.3 & 0 & & 3.50 $\pm$ 0.54 \\
		& Per49 & 03:29:12.96 & +31:18:14.31 & 27.5 & I & & 0.65 $\pm$ 0.10 \\
		B1-b & S & 03:33:21.30 & +31:07:27.40 & ... & 0 & non-clustered & 0.32 $\pm$ 0.05 \\
		& N & 03:33:21.20 & +31:07:44.20 & 17.4 & 0 & & 0.16 $\pm$ 0.05 \\
		& W & 03:33:20.30 & +31:07:21.29 & 13.9 & I & & 0.10 $\pm$ 0.02 \\
		IC348 Per8+Per55\tablefootmark{a} & Per8 & 03:44:43.94 & +32:01:36.09 & ... & 0 & non-clustered & 1.96 $\pm$ 0.30 \\
		& Per55 & 03:44:43.33 & +32:01:31.41 & 9.6 & I & & 1.58 $\pm$ 0.26 \\
		\hline
		\multicolumn{8}{c}{Close binaries}\\
		\hline
		NGC1333 IRAS1 &  & 03:28:37.00 & +31:13:27.5 & 1.908 & I & clustered & 11.00 $\pm$ 1.78 \\
		IRAS 03282+3035 &  & 03:31:21.00 & +30:45:30.0 & 0.098 & 0 & non-clustered & 1.49 $\pm$ 0.23 \\
		IRAS 03292+3039\tablefootmark{a} &  & 03:32:17.00 & +30:49:47.0 & 0.085 & 0 & non-clustered & 0.89 $\pm$ 0.14 \\
		\hline
		\multicolumn{8}{c}{Single systems}\\
		\hline
		IRAS 03271+3013 &  & 03:30:15.00 & +30:23:49.0 & ... & I & non-clustered & 1.62 $\pm$ 0.26 \\
		L1455-Per25 &  & 03:26:37.46 & +30:15:28.01 & ... & 0 & non-clustered & 1.09 $\pm$ 0.17 \\
		NGC1333 SK1 &  & 03:29:00.00 & +31:12:00.7 & ... & 0 & clustered & 0.71 $\pm$ 0.11 \\
		\hline
	\end{tabular}
    \\
    \tablefoot{\tablefoottext{a}{Only observed in \ce{13CO} with APEX.}
    \tablefoottext{b}{Clustered regions are defined to have 34 YSO~pc$^{-1}$, while non-clustered regions present 6 YSO~pc$^{-1}$, \citep{plunkett2013}.}}
	\label{tab:source}
\end{table*}

\begin{figure*}
	\centering
	\includegraphics[width=\textwidth]{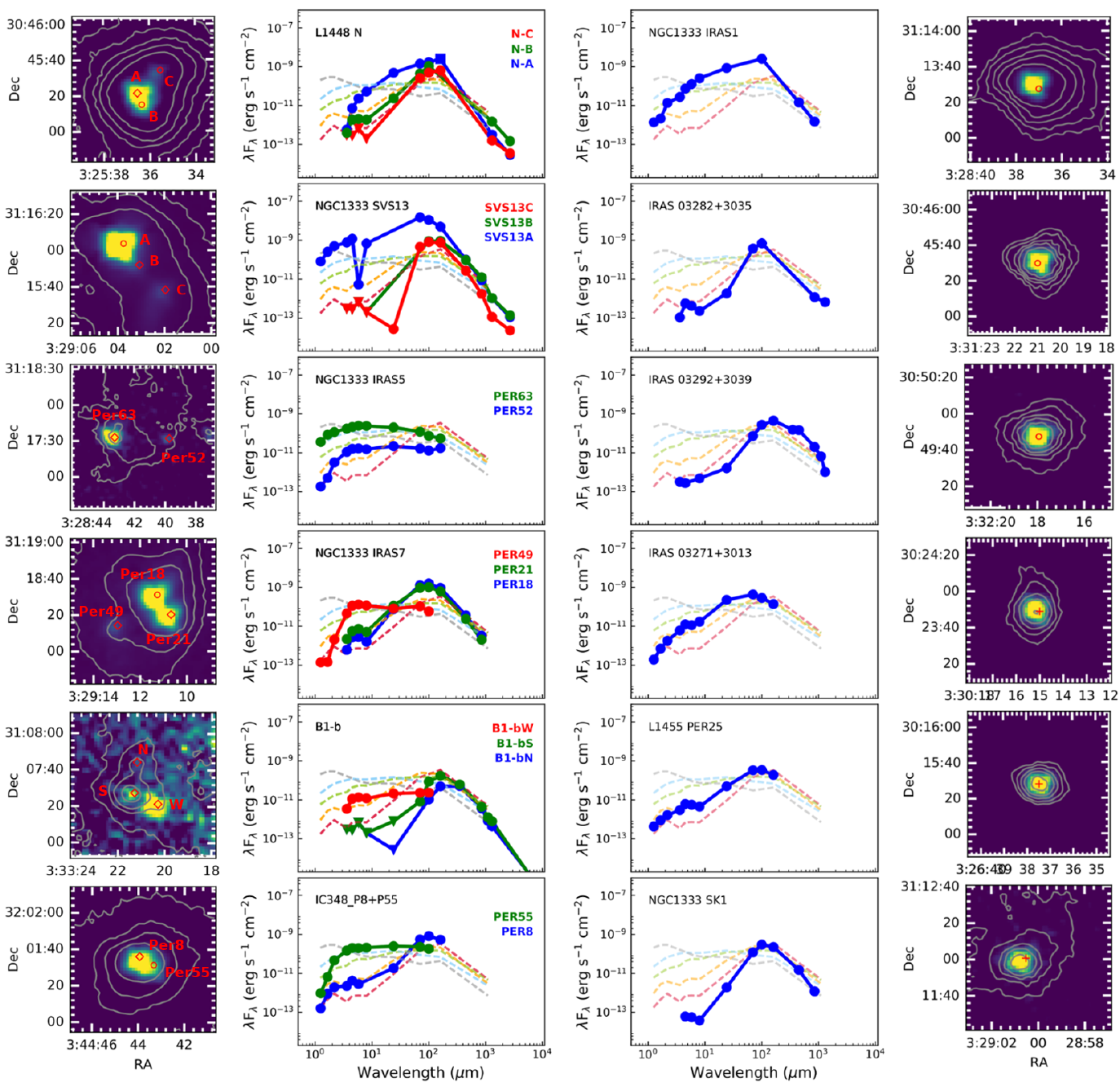}
	\caption{\textit{Herschel} PACS maps of the systems sampled in this work together with their respective SEDs. The 70 and 160 $\mu$m emission are shown in color-scale and contours, respectively. Each stamp spans a region of 80$\arcsec$ $\times$ 80$\arcsec$ and is centered on the position of the OTF maps, except for NGC1333 IRAS5 whose sources have a separation of 45.7$\arcsec$. Red symbols represent the sources of a system and the positions of the APEX single-pointing observations. Circles denote systems with additional unresolved multiplicity. Diamonds indicate single sources within multiple protostellar systems. Crosses indicate single protostars. The SEDs are overlaid on the average SEDs from Enoch et al. (2009) for reference (dashed lines), with early Class 0 (red), late Class 0, early Class I, late Class I, and Class II (gray).}
	\label{fig:sample}
\end{figure*}

\begin{table*}
	\centering
	\caption{Molecular species in this work}
	\begin{tabular}{c c c c c}
		\hline \hline
		Molecule & Transition & Frequency & E$_{\rm up}$ & log$_{10}$ A$_{\rm ij}$ \\
		& & GHz & K & \\
		\hline
		\multicolumn{5}{c}{Single pointing} \\
		\hline
		\ce{SO} & 5$_{5}$--4$_{4}$ & 215.22065 & 44.10 & -3.92 \\
		\ce{DCO+} & 3--2 & 216.11258 & 20.74 & -2.62 \\
		\ce{c-C3H2} & 3$_{3,0}$--2$_{2,1}$ & 216.27876 & 19.47 & -3.33 \\
		\ce{DCN} & 3--2 & 217.23863 & 20.85 & -4.24 \\
		\ce{c-C3H2}\tablefootmark{a} & 6--5 & 217.82215 & 38.61 & -3.23 \\
		\ce{c-C3H2} & 5$_{1,4}$--4$_{2,3}$ & 217.94005 & 35.42 & -3.35 \\
		\ce{p-H2CO} & 3$_{0,3}$--2$_{0,2}$ & 218.22219 & 20.96 & -3.55 \\
		\ce{CH3OH} & 4$_{2,2}$--3$_{1,2}$ & 218.44005 & 45.45 & -4.33 \\
		\ce{p-H2CO} & 3$_{2,2}$--2$_{2,1}$ & 218.47563 & 68.09 & -3.80 \\
		\ce{p-H2CO} & 3$_{2,1}$--2$_{2,0}$ & 218.76007 & 68.11 & -3.80 \\
		\ce{C2H} & 4--3 J=9/2--7/2 F=5--4 & 349.33771 & 41.91 & -3.88 \\
		\ce{C2H} & 4--3 J=9/2--7/2 F=4--3 & 349.33899 & 41.91 & -3.89 \\
		\ce{C2H} & 4--3 J=7/2--5/2 F=4--3 & 349.39927 & 41.93 & -3.90 \\
		\ce{C2H} & 4--3 J=7/2--5/2 F=3--2 & 349.40067 & 41.93 & -3.92 \\
		\ce{o-H2CO} & 5$_{1,5}$--4$_{1,4}$ & 351.76864 & 62.45 & -2.92 \\
		\ce{DCO+} & 5--4 & 360.16978 & 51.86 & -2.42 \\
		\ce{DCN} & 5--4 & 362.04575 & 52.12 & -3.13 \\
		\ce{HNC} & 4--3 & 362.63030 & 43.51 & -2.64 \\
		\ce{p-H2CO} & 5$_{0,5}$--4$_{0,4}$ & 362.73602 & 52.31 & -2.86 \\
		\hline
		\multicolumn{5}{c}{APEX OTF maps}\\
		\hline
		\ce{^{13}CO} & 4--3 & 440.76517 & 52.9 & -5.27 \\
		\ce{^{13}CO} & 6--5 & 661.06728 & 111.1 & -4.73 \\
        \hline
        \multicolumn{5}{c}{JCMT and Herschel HIFI}\\
        \hline
   		\ce{^{13}CO} & 3--2 & 330.58797 & 31.7 & -5.66 \\
   		\ce{^{13}CO} & 10--9 & 1101.34960 & 290.8 & -4.05 \\
		\hline
	\end{tabular}
	\\
	\tablefoot{\tablefoottext{a}{Contains both ortho- and para forms.}}
	\tablebib{All rest frequencies were taken from the Cologne Database for Molecular Spectroscopy (CDMS; \citealt{CDMS_2016}).
		The \ce{SO} entry is based on \citet{so1976}. The \ce{DCO+} entries are based on \citet{DCO+_rot_2005}. The \ce{c-C3H2} 
		entry was based on \citet{c-C3H2_isos_rot_1987} with transition frequencies 
		important for our survey from \citet{c-C3H2_rot_1986} and from \citet{c-C3H2_isos_rot_2012}. 
		The \ce{DCN} entries are based on \citet{hnc1993} and \citet{dcn2004}.
		The entry for \ce{CH3OH} is based on the line list from \citet{ch3oh1997}.
		The \ce{H2CO} entries are based on experimental data from \citet{h2co1996}.
		The \ce{C2H} entry is based on \citet{CCH_obs_par_2009} with additional important data 
		from \citet{CCH_rot_2000} and \citet{CCH_rot1981}. 
		and on \citet{cations_rot_2012}, respectively. 
		The entry for \ce{HNC} is based on \citet{hnc1993}.
		The \ce{^{13}CO} entries are based on \citet{13co2000}.}
	\label{tab:lines}
\end{table*}

\section{System sample}
\label{sec:sample}

The Perseus molecular cloud ($d = 235$ pc) large-scale structure has been well studied in continuum (e.g. \citealt{hatchell2005,enoch2006,chen2016,pokhrel2018}) and molecular line emission (e.g. \citealt{arce2010,curtis2010a,curtis2010b,curtis2011,walker2014,hacar2017}).
The small scale structure of the region has been studied through the characterization of individual systems (e.g. \citealt{kwon2006,enoch2009,mottram2013,hirano2014,dauw2016}).
The evolutionary classification of the protostars in the region has been carefully studied through molecular line and continuum observations \citep{enoch2009,carney2016,mottram2017}.
Recently, \citet{tobin2016} conducted an unbiased 8 mm survey of all protostars in Perseus down to 15 AU separation with the Karl G. Jansky Very Large Array (VLA), thus characterizing the multiplicity of the star forming region.

The evolutionary stages of each source in embedded wide multiple protostellar systems in Perseus have been characterized through the construction of spectral energy distributions (SEDs) and the parameters derived from the SEDs including \textit{Herschel} Space Observatory PACS maps \citep{murillo2016}. This provides information on the coevality of wide multiple protostellar systems and can help to understand how these systems are formed \citep{murillo2016}.
Further examination of coevality, core structure and protostar distribution was done by \citet{sadavoy2017}, studying possible formation mechanisms. 
Dust emission observed with the VLA toward several disk-candidate embedded protostellar systems was examined by \citet{seguracox2016}, and the dust continuum was found to present disk-shaped structures.
In addition, envelopes and outflows driven by protostars in Perseus have been studied both at scales larger than 4000 AU \citep{davis2008,curtis2010a,curtis2010b,arce2010,mottram2013,karska2014,yildiz2015,mottram2017} and below 2000 AU \citep{persson2012,plunkett2013,maret2014,lee2016}.
Together, previous work provides an extensive database of information about the protostars in the Perseus molecular cloud.

For this study, a sample of 12 low-mass protostellar systems in Perseus were selected from the work of \citet{tobin2016} and \citet{murillo2016}.
The sample is listed in Table~\ref{tab:source}, along with coordinates, source separations, type of region where they are located and the bolometric luminosity $L_{\rm bol}$ calculated from the SEDs.
Figure~\ref{fig:sample} shows \textit{Hersechel} PACS thermal continuum mini-maps of the sample obtained from the Gould Belt Survey \citep{andre2010}, along with the constructed SEDs from \citet{murillo2016}.
The selected systems are young embedded protostars in the Class 0 and I evolutionary stages.
Both single and multiple (i.e., binary and higher order multiples) protostellar systems are included, with the multiple systems spanning a range of separations from $\sim$0.1$\arcsec$ to 46$\arcsec$ ($\sim$23.5 AU to 11000 AU).
Thus, both close and wide multiple protostellar systems are considered in this study.
Finally, the systems are located in both clustered (NGC1333) and non-clustered regions (L1448, L1455 and B1).
Selecting systems from both clustered and non-clustered regions (34 and 6 YSO pc$^{-2}$, respectively; \citealt{plunkett2013}) allows the impact of external heating on the measured gas temperatures to be assessed.

This sample then allows the evolutionary stage, multiplicity, and region to be compared with temperature, both from the UV-heated gas and the envelope gas temperature.
Since the timescale for protostellar evolution (of order a few $\times$10$^{5}$ yrs, \citealt{evans2009,mottram2011,heiderman2015,carney2016}) is considerably longer than can be observed in human lifespan, the evolution of the protostellar temperature structure cannot be directly observed. Hence, systems in the Class 0 and I evolutionary stages need to be compared, as well as single and multiple protostellar systems.
Evolutionary classes are determined based on the shape of the SED, derived parameters such as bolometric temperature $T_{\rm bol}$, and the structure of the system (e.g. envelope, outflow opening angle).
Thus, the temperature-multiplicity-age relation can be studied, which can in turn provide constraints for hydrodynamical models with radiative feedback.

\section{Observations}
\label{sec:obs}
\subsection{Single pointing}
\label{subsec:points}
APEX observations of 10 out of 12 systems in our sample were carried out with the Swedish Heterodyne Facility Instrument (SHeFI; \citealt{nystrom2009}) in position switching mode.
The APEX-1 band was used for observations on 1 December 2016 with a precipitable water vapor (PWV) $\sim$ 1.6 mm (O-098.F-9320B.2016, NL GTO time) using one spectral setting with the central frequency set to 217.11258 GHz and a bandwidth of 4 GHz.
This setting targeted the molecules \ce{DCO+}, \ce{DCN}, \ce{c-C3H2} and \ce{H2CO}.
In addition, transitions of \ce{SO} and \ce{CH3OH} were detected.
Typical noise levels for the observations ranged between 15 to 70 mK for a channel width of 0.4 km s$^{-1}$ and a HPBW of 28.7$\arcsec$. 
The beam efficiency $\eta_{\rm mb}$ for observations at 230 GHz is 0.75.

APEX-2 band observations were carried out from 7 to 12 July 2017 with PWV between 0.37 and 1.5 mm (M-099.F-9516C-2017) using two spectral settings with central frequencies of 350.33746 and 361.16978 GHz, and bandwidth of 4 GHz.
The molecules targeted were \ce{C2H}, \ce{DCO+}, \ce{DCN} and \ce{H2CO}. \ce{HNC} was also detected.
Typical noise levels range from 20 to 100 mK for a channel width of 0.4 km s$^{-1}$, a HPBW of 18$\arcsec$ and $\eta_{\rm mb}$ = 0.73.
For both bands, calibration uncertainties are on the order of 20\%.
The molecules targeted in these observations probe the cold (\ce{DCO+}, \ce{H2CO} 218.222 GHz) and warm (\ce{DCN}, \ce{c-C3H2}, \ce{C2H}, \ce{H2CO}) gas of the envelope at scales of 7000 AU and a gas temperature range of 10 to 100 K (Table~\ref{tab:lines}).

\subsection{OTF maps}
\label{subsec:maps}
On-the-fly (OTF) maps of all 12 systems were obtained with two instruments: CHAMP+ \citep{kasemann2006}, 
and SEPIA B9 \citep{baryshev2015,belitsky2017}, in order to observe \ce{^{13}CO} 6--5. 
\ce{^{13}CO} 6--5 is a particularly useful tracer of UV heated gas, in contrast to \ce{^{12}CO} 6--5 \citep{yildiz2012,vankempen2010}.

CHAMP+ was used to observe three systems: NGC1333 IRAS7, IC348 Per8+Per55 and IRAS 03292+3029, with a spectral set-up targeting \ce{^{13}CO} 6--5 (661.06728 GHz) and a beam of 9.4$\arcsec$ (HPBW). 
Maps of 45$\arcsec~\times$~45$\arcsec$ were centered on the target system in the case of IC348 Per8+Per55 and IRAS 03292+3029. In the case of NGC1333 IRAS7, the maps were centered at a position equidistant from all sources. 
Observations took place in two epochs, from 26 August to 19 September 2014 (M-094.F-0006.2014), and 2 to 13 of August 2015 (M-095.F-0023.2014).

SEPIA B9 maps (45$\arcsec~\times$~45$\arcsec$) were made for the remaining 9 systems in our sample, with the spectral set-up targeting \ce{^{13}CO} 6--5 (661.06728 GHz) and a beam of 9.4$\arcsec$ (HPBW).
Two maps were made for each of the systems NGC1333 SVS13 and NGC1333 IRAS5, given the separation between the sources. For NGC1333 SVS13, the maps were centered on the A and C sources.
Observations took place from 13 August to 25 November 2016 (O-098.F-9320A.2016), with an additional science verification observation of B1-B on 29 July 2016 (E-097.F-9810A.2016).

The systems IRAS 03282+3035 and IRAS 03292+3029 were further observed using FLASH \citep{heyminck2006} with a spectral set-up targeting \ce{^{13}CO} 4--3 (440.76517 GHz), since observations of \ce{^{13}CO} 3--2 were not available with JCMT for these two systems. 
Observations were carried out on 26 August 2014 (M-094.F-0006.2014).

In order to compare the observations from FLASH, CHAMP+ and SEPIA B9 with the Herschel HIFI \ce{^{13}CO} 10--9 observations (RMS noise: 0.03 K; \citealt{mottram2017}), the data were averaged within a box of approximately 19.3$\arcsec$, the HPBW of the HIFI observations, centered on the position of the HIFI observations.
The FLASH observations provide RMS noise of 0.44 and 0.75 K for IRAS 03292+3029 and IRAS 03282+3035, respectively, for a channel width of 0.4 km~s$^{-1}$.
For the CHAMP+ observations, typical RMS noise is between 0.03 to 0.08 K for a channel width of 0.4 km~s$^{-1}$.
For the SEPIA B9 observations the typical RMS noise level is between 0.1 to 0.2 K for a channel width of 0.4 km~s$^{-1}$.
The main beam efficiencies $\eta_{\rm mb}$ being used are 0.60 for 440 GHz, and 0.56 for 660 GHz.
Typical calibration uncertainties are about 10 to 20\%.

\section{Results}
\label{sec:results}

\begin{figure*}
	\centering
	\includegraphics[width=0.98\textwidth]{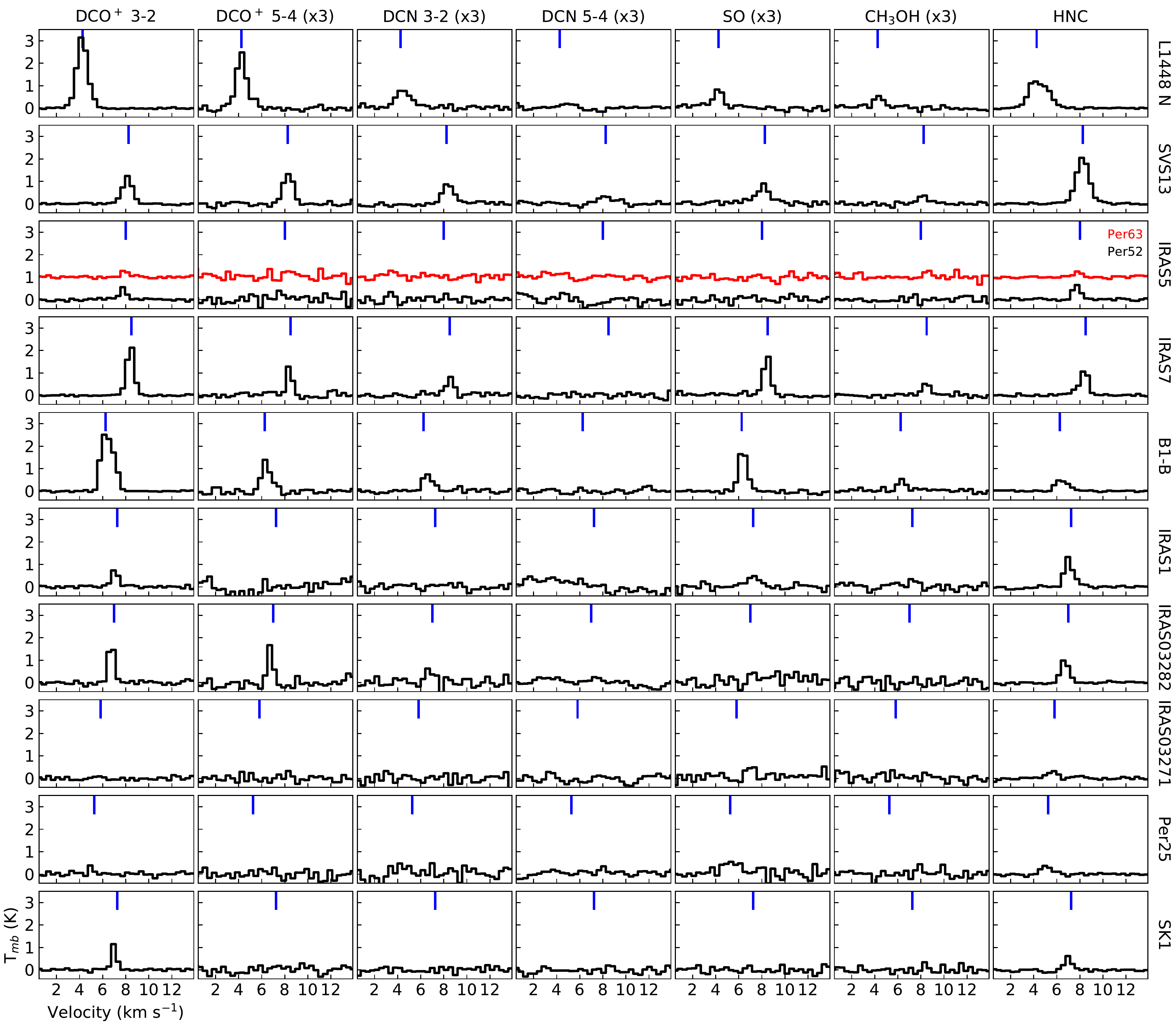}
	\caption{APEX single pointing observations of \ce{DCO+}, \ce{DCN}, \ce{SO}, \ce{CH3OH} and \ce{HNC}. Except for \ce{DCO+} 3--2 and \ce{HNC}, the spectra are multiplied by a factor of 3 to make the features more prominent. The spectra for the wide multiple systems are averaged for all the sources, except for NGC1333 IRAS5. The vertical blue line marks the systemic velocity.}
	\label{fig:dco_dcn_so_ch3oh_hnc}
\end{figure*}

\begin{figure*}
	\centering
	\includegraphics[width=\textwidth]{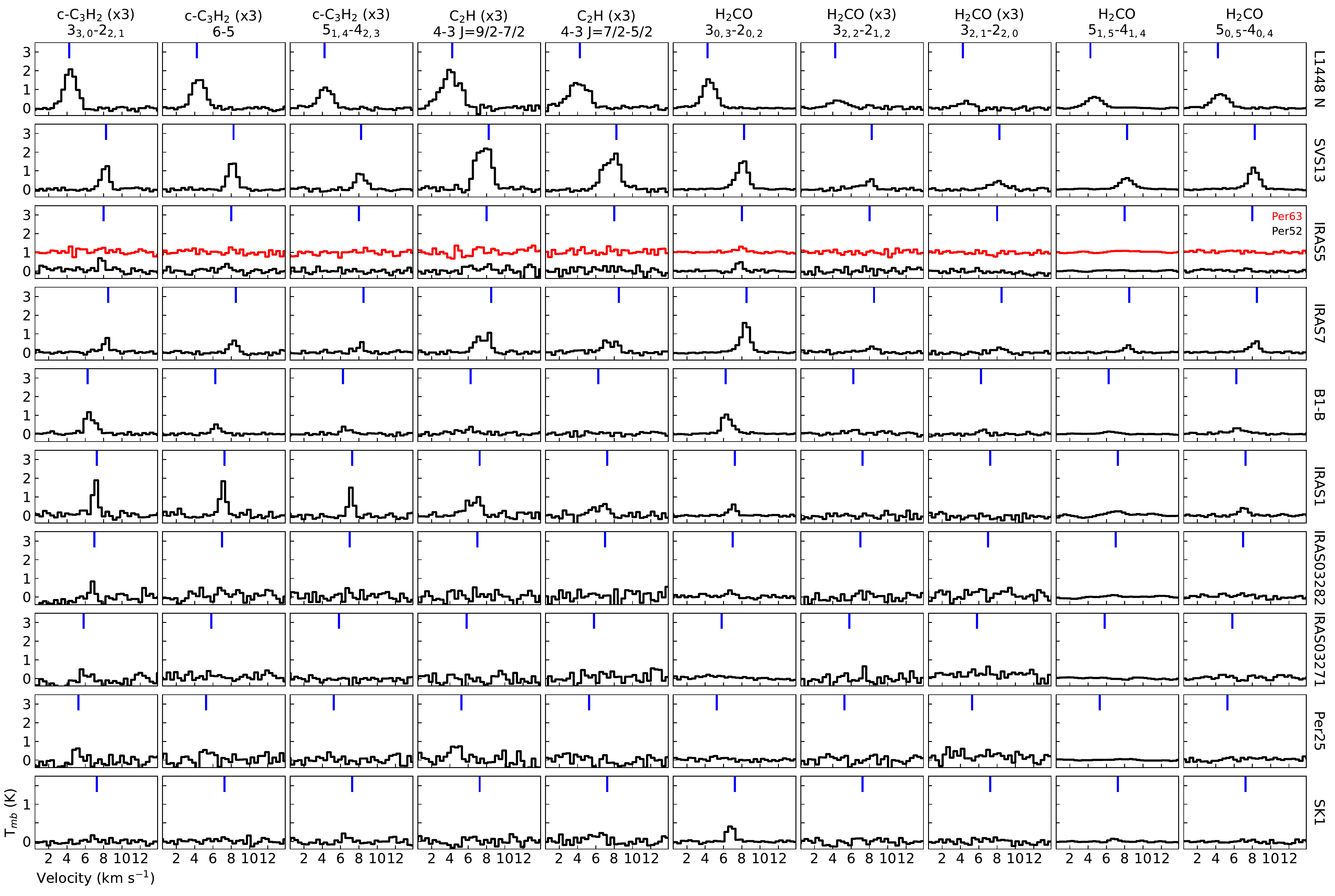}
	\caption{Same as Fig.~\ref{fig:dco_dcn_so_ch3oh_hnc} but for \ce{c-C3H2}, \ce{C2H} and \ce{H2CO}.}
	\label{fig:c3h2_c2h_h2co}
\end{figure*}

\subsection{Cold and warm gas}
\label{subsec:coldwarm}
The observed molecular lines trace the cold and warmer envelope gas of the systems in our sample.
For the multiple protostellar systems, single pointing observations for each source were taken. However, the beams of the observations (28.7$\arcsec$ and 18$\arcsec$) are comparable to the source separations, and the spectra of the individual sources in a wide multiple system are similar.
As an example, the spectra of the individual sources for NGC1333 IRAS7 are shown in Fig.~\ref{fig:IRAS7}.
Thus the spectra of all three sources in the wide multiple systems are averaged together, except for NGC1333 IRAS5 which has a separation of 45.7$\arcsec$ and is treated in this work as two separate single protostars Per63 and Per52.
The results of the averaged spectra are discussed and analyzed in this work.
Figures~\ref{fig:dco_dcn_so_ch3oh_hnc} and \ref{fig:c3h2_c2h_h2co} show the resulting spectra of the observed sample.
Observed line peak temperatures, RMS noise, line widths and integrated fluxes are listed in Tables~\ref{tab:dcodcn} to \ref{tab:h2co_5-4}.

\ce{HNC} was detected toward all the systems in the sample, however, the other observed molecules were not detected in the envelope of every system.
Three out of the five wide multiple protostellar systems show detections with signal-to-noise (S/N) above 3 in 0.4 km~s$^{-1}$ channels for all targeted molecular lines.
B1-b, also a wide multiple system, shows above 3$\sigma$ detections in all lines except \ce{H2CO} 3$_{2,2}$--2$_{1,2}$, \ce{DCN} 5--4 and \ce{C2H} 4--3 J=7/2--5/2.
\ce{SO} and methanol (\ce{CH3OH}) were detected toward four wide multiple systems. 
IRAS5 Per63 has detections above 3$\sigma$ only in \ce{DCO+} 3--2, \ce{H2CO} 3$_{0,3}$--2$_{0,2}$ and \ce{H2CO} 5$_{0,5}$--4$_{0,4}$. 
In contrast, IRAS5 Per52 presents detections in \ce{DCO+} 3--2, \ce{c-C3H2} 3--2 and 6--5, and \ce{H2CO} 3$_{0,3}$--2$_{0,2}$.

The close binary system NGC1333 IRAS 1 presents emission with S/N$>$3 in \ce{SO}, \ce{DCO+} 3--2, all transitions of the warm molecules \ce{c-C3H2} and \ce{C2H}, as well as \ce{H2CO} 3$_{0,3}$--2$_{0,2}$.
In contrast, IRAS 03282+3035 presents emission only in both transitions of \ce{DCO+}, \ce{DCN} 3--2 and \ce{H2CO} 3$_{0,3}$--2$_{0,2}$.

The single protostellar systems show little to no emission.
NGC1333 SK1 presents emission above 3$\sigma$ in \ce{DCO+} 3--2, \ce{H2CO} 3$_{0,3}$--2$_{0,2}$ and \ce{H2CO} 5$_{0,5}$--4$_{0,4}$; while L1455-Per25 shows only \ce{DCO+} 3--2 and \ce{HNC} emission.
The other single protostar, IRAS 03271+3013, presented no detection beyond \ce{HNC}. 
This might be due to the protostar being a Class I protostar, however, NGC1333 IRAS 1 and both sources of NGC1333 IRAS5 are also Class I objects but show more line detections than IRAS 03271+3013.
In summary, wide multiple systems generally present more line emission than single protostars, most likely related to the higher envelope mass, and hence larger molecular column density.

\ce{DCO+} presents strong emission towards most systems. Since \ce{DCO+} is formed from the reaction of \ce{H2D+} $+$ \ce{CO}, with \ce{H2D+} greatly enhanced at low temperatures due to the freeze-out of CO ($<$30 K), \ce{DCO+} has been found to be a good tracer of cold gas \citep{jorgensen2005,mathews2013,murillo2015}.
It should be noted, however, that \ce{CO} freezes out onto the dust grains at densities above 10$^{4}$ $\sim$ 10$^{5}$ cm$^{-3}$ (e.g., \citealt{caselli1999,bergin2002,jorgensen2005}), thus at these densities \ce{DCO+} may be dependent on density as well as temperature.
The low-lying transition of \ce{H2CO} 3$_{0,3}$--2$_{0,2}$ is the strongest among the five transitions, with peaks a factor of $\sim$10 higher than the higher-lying \ce{H2CO} transitions. The peaks of the \ce{H2CO} 5--4 transitions appear stronger than the higher-lying 3--2 transitions, this is due to the smaller HPBW of the APEX-2 observations (18$\arcsec$).
\ce{DCN} is a warm gas tracer, which can be formed and fractionated through a higher temperature route starting with \ce{CH2D+} (e.g., \citealt{favre2015}). Similarly to \ce{DCN}, the higher-lying transitions of \ce{H2CO} also trace warm gas.
The weak emission of \ce{DCN} and the higher-lying transitions of \ce{H2CO} would suggest that the envelope gas is relatively cold and currently not being strongly heated.
The molecules \ce{c-C3H2} and \ce{C2H} trace the warm UV-irradiated gas (e.g., \citealt{nagy2015,guzman2015}), most likely located along the outflow cavity (e.g., \citealt{fontani2012,jorgensen2013,murillo2018}).
The peak intensities of \ce{c-C3H2} vary by less than a factor of 3 among all three observed transitions.
\ce{C2H} is detected in both spin doubling transitions with each transition showing a characteristic double hyperfine structure pattern.
Methanol (\ce{CH3OH}) and \ce{SO} are mainly formed on grain surfaces, and either are sputtered off the grains by shocks (e.g., \citealt{buckle2002,burkhardt2016}) or sublimated into the gas phase in hot, dense regions of the outflow (e.g., \citealt{vandertak2000,palau2017}).

The different systemic velocities of each region within Perseus are reflected in the observed spectra (Fig.~\ref{fig:dco_dcn_so_ch3oh_hnc} and \ref{fig:c3h2_c2h_h2co}).
For the systems located in NGC1333, there is also a slight difference in the systemic velocity between the systems located closer to the cluster center (NGC1333 SVS13, NGC1333 IRAS7 and NGC1333 IRAS5; $v_{\rm LSR}~=$ 8.0--8.5 km~s$^{-1}$) and those located in the outer part of the cluster (NGC1333 IRAS1 and NGC1333 SK1; $v_{\rm LSR}~\sim$ 7.3 km~s$^{-1}$).

\subsection{\ce{^{13}CO} maps}
\label{subsec:uv}
The results from the \ce{^{13}CO} maps are described in this section. The spectra extracted from the maps are shown in Fig.~\ref{fig:13co}, along with the spectra from JCMT and \textit{Herschel} HIFI observations, for comparison.
Spectra from the JCMT and APEX observations are smoothed to the beam of the HIFI observations.
Observed line peak temperatures, RMS noise, line widths and integrated fluxes are listed in Table~\ref{tab:13co}.

\ce{^{13}CO} 4--3 was observed toward the close binary systems IRAS 03282+3035 and IRAS 03292+3039, with emission being detected only toward IRAS 03292+3039 (S/N$\sim$7$\sigma$). 
IRAS 03282+3035 does not show emission at the natural resolution of the FLASH observations ($\sigma\sim$0.5 K), nor smoothed to the \textit{Herschel} HIFI observations HPBW ~=~19.25$\arcsec$ ($\sigma\sim$1 K).

\ce{^{13}CO} 6--5 was mainly detected toward the wide multiple systems and one close binary system (Fig.~\ref{fig:13co}). 
NGC1333 IRAS7 and IC348 Per8+Per55, observed with CHAMP+, present strong emission.
For NGC1333 IRAS7, the bulk of the emission (S/N~=~24$\sigma$) is located between Per18 and Per21, with little emission toward Per49, while IC348 Per8+Per55 shows centrally concentrated emission with S/N~=~42$\sigma$.
IRAS 02393+3039 did not show any significant line emission in the CHAMP+ observations with a noise of 0.12 K.

Observations with SEPIA B9 detected \ce{^{13}CO} 6--5 toward L1448N, NGC1333 SVS13 and NGC1333 IRAS1.
From these systems, NGC1333 SVS13 presents the strongest emission, peaking at $\sim$4.7 K (average of the central position of the maps centered on sources A and C). L1448N presents weak emission, despite presenting relatively strong line detections in all the other molecules observed with APEX-1 and APEX-2 (Fig.~\ref{fig:dco_dcn_so_ch3oh_hnc} and \ref{fig:c3h2_c2h_h2co}). However, the weak \ce{^{13}CO} 6--5 emission is consistent with the observations of \ce{^{13}CO} using the JCMT and \textit{Herschel} HIFI (Table~\ref{tab:13co} and Fig.~\ref{fig:13co}).

The SEPIA B9 observations show considerably more noise by a factor of 3 higher than those of the CHAMP+ observations.
Considering the signal-to-noise ratio of the detected emission, however, the non-detections are not due to the higher noise level of the SEPIA B9 observations, but most likely from the compact \ce{^{13}CO} 6--5 emission toward these systems, which is diluted in the larger beam.

\begin{table*}
	\centering
	\caption{Source parameters}
	\begin{tabular}{c c c c c}
		\hline \hline
		System & 850$\mu$m peak intensity & $M_{\rm env}$ & $L_{\rm bol}$ & $M_{\rm env}$ / $L_{\rm bol}$ \\
		& Jy~beam$^{-1}$ & M$_{\odot}$ & L$_{\odot}$ & M$_{\odot}$ / L$_{\odot}$ \\
		\hline
		\multicolumn{5}{c}{Wide multiples} \\
		\hline 
		L1448N 	&	5.69	&	2.920	&	13.08	&	0.22	\\
		SVS13 	&	4.63	&	1.023	&	121.07	&	0.01	\\
		Per63	&	0.33	&	0.215	&	1.38	&	0.16	\\
		Per52	&	0.04	&	0.041	&	0.12	&	0.34	\\
		IRAS7	&	1.76	&	0.820	&	8.92	&	0.09	\\
		B1-b 	&	2.33	&	3.052	&	0.59	&	5.17	\\
		Per8+Per55	&	0.88	&	0.506	&	3.38	&	0.15	\\
		\hline
		\multicolumn{5}{c}{Close binaries} \\
		\hline
		IRAS1 	&	0.86	&	0.322	&	11	&	0.03	\\
		IRAS 03282	&	1.17	&	0.957	&	1.49	&	0.64	\\
		IRAS 03292	&	2.43	&	2.768	&	0.89	&	3.11	\\
		\hline
		\multicolumn{5}{c}{Singles} \\
		\hline 
		IRAS03271	&	0.24	&	0.139	&	1.62	&	0.09	\\
		Per25	&	0.35	&	0.252	&	1.09	&	0.23	\\
		SK1	&	0.33	&	0.274	&	0.71	&	0.39	\\
		\hline
	\end{tabular}
	\label{tab:mass}
\end{table*}

\begin{figure*}
	\centering
	\includegraphics[width=0.98\textwidth]{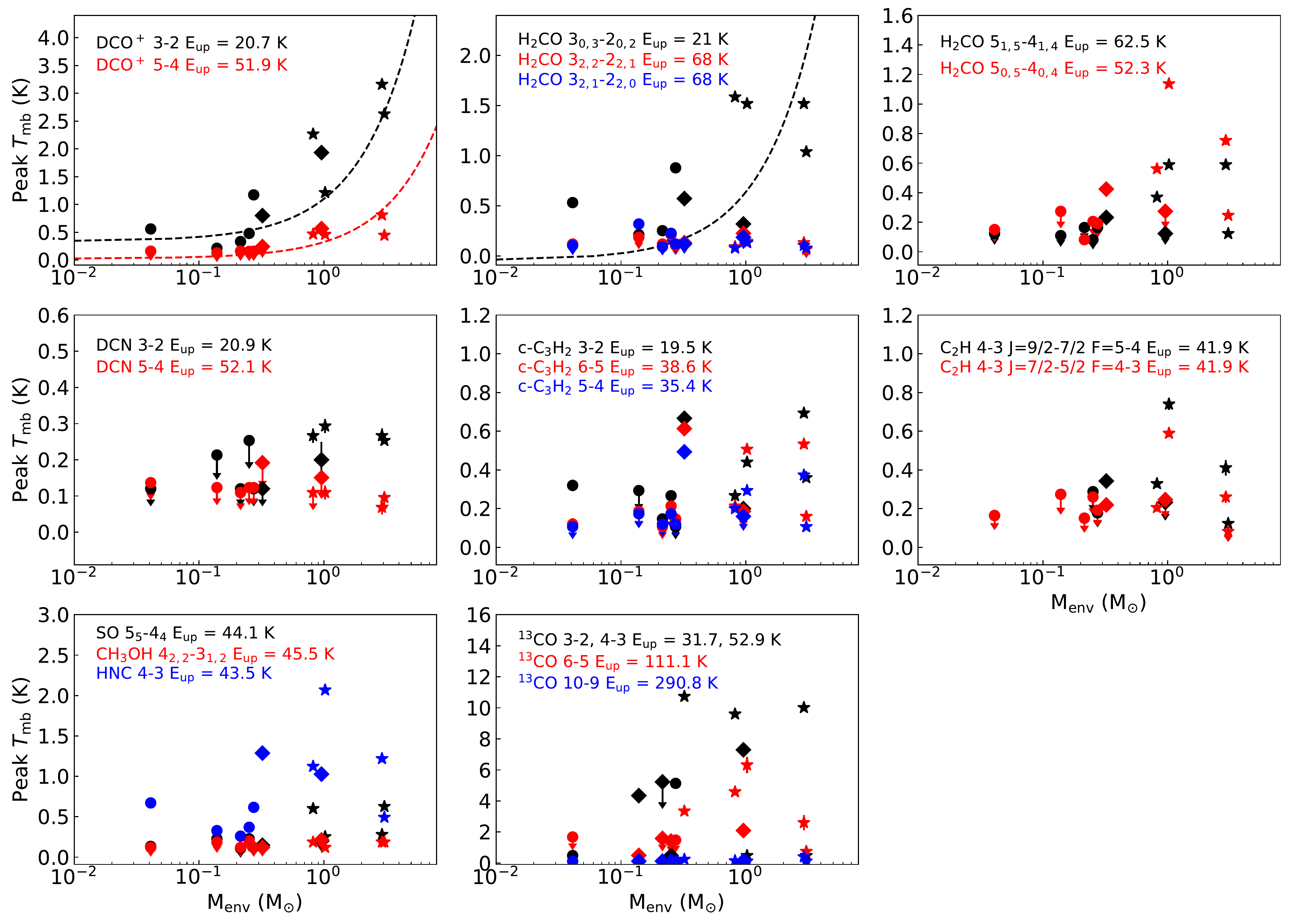}
	\caption{Peak intensities of the observed molecular lines compared to the envelope mass (M$_{\odot}$) of each systems. The dashed lines are linear fits to the data for the cases where a correlation is found. Circles, diamonds and stars show single, close binary and wide multiple protostellar systems, respectively. Note that the more massive envelopes show an increase in the peak intensities of \ce{DCO+} and the low-lying transition of \ce{H2CO}, which trace cold gas in the envelope. Molecules tracing warm gas have similar peak intensities regardless of envelope mass.}
	\label{fig:masspeak}
\end{figure*}

\begin{figure*}
	\centering
	\includegraphics[width=0.98\textwidth]{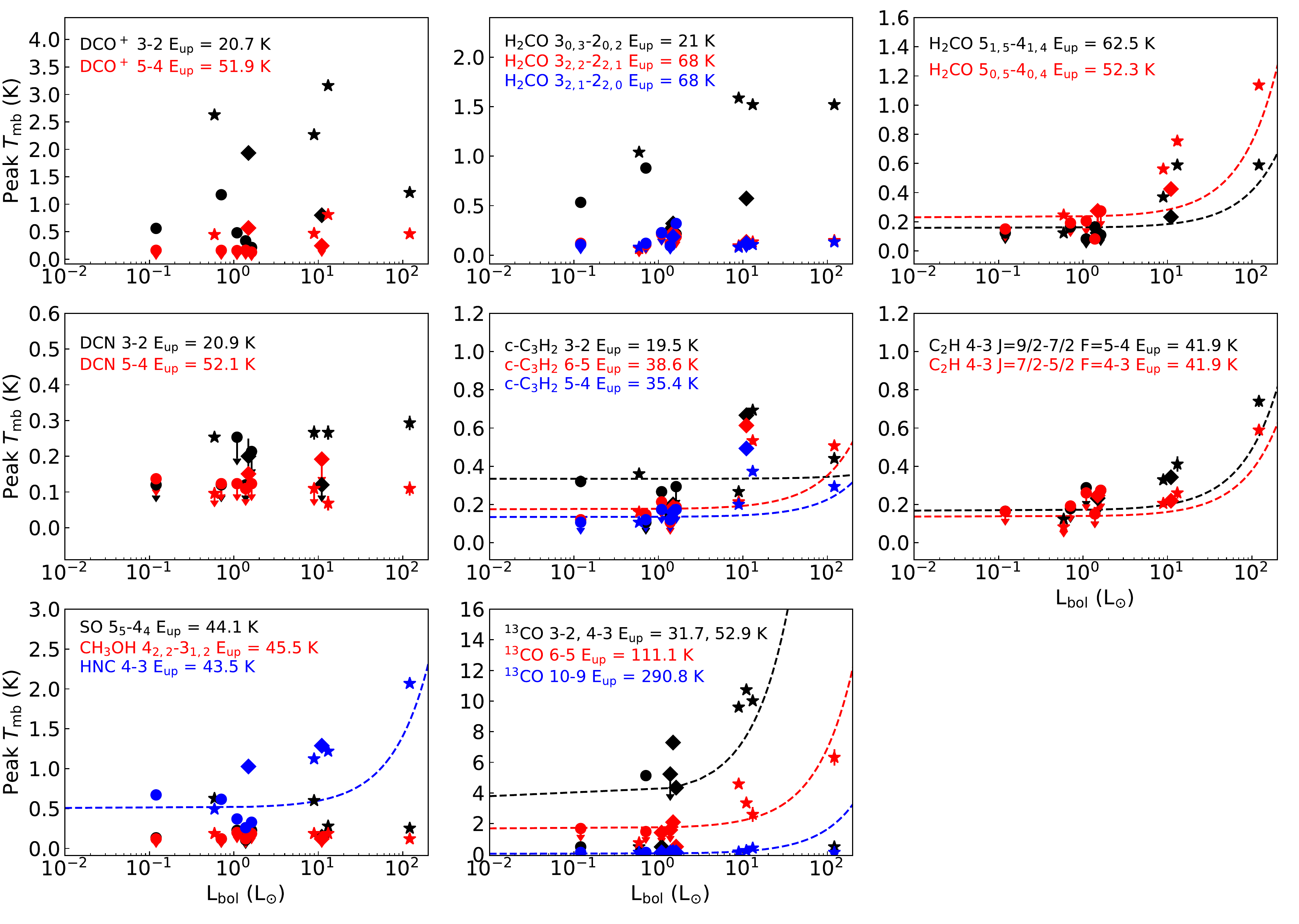}
	\caption{Peak intensities of the observed molecular lines compared to the bolometric luminosity $L_{\rm bol}$ (L$_{\odot}$) of each system. The dashed lines are linear fits to the data for the cases where a correlation is found. Circles, diamonds and stars show single, close binary and wide multiple protostellar systems, respectively. \ce{C2H} and \ce{c-C3H2} show somewhat higher peak intensities in systems with relatively higher luminosities.}
	\label{fig:Lbolpeak}
\end{figure*}

\section{Analysis}
\label{sec:analysis}

\subsection{Line emission and system parameters}
\label{subsec:peakparam}

The observed molecular line emission is compared to system luminosity and envelope mass in this section.
Bolometric luminosity $L_{\rm bol}$ is obtained from the SEDs of the observed systems, with $L_{\rm bol}$ for the wide multiple systems derived from the combined SEDs \citep{murillo2016}. 
Envelope mass $M_{\rm env}$, listed in Table~\ref{tab:mass}, is calculated from the 850 $\mu$m peak intensity $S_{\rm 850\mu m}$, $L_{\rm bol}$ and distance $d$ using the formula from \citet{jorgensen2009} expressed as:
\begin{equation}
M_{\rm env} = 0.44 M_{\odot}~\left(\frac{L_{\rm bol}}{1~L_{\odot}}\right)^{-0.36}~\left(\frac{S_{850\mu m}}{1~{\rm Jy~beam}^{-1}}\right)^{1.2}~\left(\frac{d}{125~{\rm pc}}\right)^{1.2}
\end{equation}
which takes into account that more luminous systems have somewhat higher dust temperatures.
The relation assumes optically thin emission and typical dust-to-gas ratio of 1:100, and is derived from the power-law relations that arise between envelope mass obtained from radiative transfer models, and the observed peak flux, and luminosity. The power-law relations are $M_{\rm env} \propto S_{850}^{1.2}$ and $M_{\rm env}/S_{850} \propto L_{\rm bol}^{0.36}$.

The 850$\mu$m peak intensity used in this work is measured from the COMPLETE survey map of Perseus taken with SCUBA on the JCMT \citep{kirk2006}, which has a beam of 15$\arcsec$.
The peak intensity was measured in a circular region of 28.7$\arcsec$ (HPBW of the APEX-1 observations) centered on each system. 
For the wide multiple systems, the total flux of all sources is used.
In addition, the observed line emission is compared to the ratio of envelope mass to bolometric luminosity, ($M_{\rm env} / L_{\rm bol}$), listed in Table~\ref{tab:mass}.
This ratio gives insight into the amount of mass heated by the luminosity of the protostellar sources within each system. Because younger systems are expected to have more mass in their envelopes, higher ratios are expected to correspond to younger sources \citep{bontemps1996}. This holds true for the close binary and single systems in our sample. But it is more difficult to disentangle for the wide multiple systems, since the individual sources present different evolutionary stages (Fig.~\ref{fig:sample} and Table~\ref{tab:source}).

System type (wide multiple, close binary and single protostar) is indicated in the plots with different symbols in order to determine if there is any relation with respect to line emission or system parameters.
For systems with molecular line non-detections, the upper limits in the plots are placed at 3$\sigma$.
A linear fit to the data is used to identify trends between the observed line emission peaks and system parameters in cases where there is a significant correlation.
Further statistical analysis is treated in Sec.~\ref{subsec:stats}.

Peak antenna temperatures are compared with the envelope mass in Fig.~\ref{fig:masspeak}.
The peak intensities of \ce{H2CO} 3$_{0,3}$--2$_{0,2}$, and \ce{DCO+} 3--2 increase with envelope mass.
Wide multiple systems have larger envelope masses than the close binaries and single protostars, with the exception of NGC1333 IRAS5, where Per63 and Per52 have envelope masses comparable to single protostars.
This can be interpreted as wide multiple systems having more massive reservoirs of cold gas compared to close binaries and single protostellar systems.
On the other hand, the warm gas being traced by \ce{DCN} and the two higher-lying transitions of \ce{H2CO} 3--2 do not show a dependency on the envelope mass, degree of multiplicity or region type. 
Instead the line peaks are practically constant with envelope mass.
Methanol, \ce{SO} and the three transitions of \ce{c-C3H2} do not show particular dependency on envelope mass either.
For \ce{^{13}CO} 3--2, there appears to be a slight correlation between envelope mass and peak antenna temperature, but there are not enough data points to be certain. The 6--5 and 10--9 transitions also do not show a correlation to envelope mass.

Figure~\ref{fig:Lbolpeak} shows the observed line peak antenna temperatures compared with $L_{\rm bol}$.
The warm molecule \ce{C2H} appears to be associated with $L_{\rm bol}$, as well as the \ce{c-C3H2} 6--5 and 5--4 transitions.
Since these molecules are generally formed in irradiated regions \citep{fontani2012,jorgensen2013,nagy2015,guzman2015}, the correlation between \ce{C2H} and \ce{c-C3H2}, and $L_{\rm bol}$ can be explained by the outflow cavity being irradiated by the central protostar, as was also found for O[I] and H2O (e.g. \citealt{mottram2017}).
Thus, the more luminous the central protostar, the deeper the outflow cavity is irradiated and more \ce{c-C3H2} and \ce{C2H} is produced.
The two transitions of \ce{H2CO} 5--4 show a correlation to bolometric luminosity, whereas the higher lying transitions of \ce{H2CO} 3--2 do not, despite the similar upper energy levels ($E_{\rm up}$ = 52--68 K). The reason for this discrepancy could be the difference in beam size from the observations. The \ce{H2CO} 3--2 transition was observed with a beam of 28.7$\arcsec$, compared to 18$\arcsec$ for the 5--4 transition. This suggests that the \ce{H2CO} 5--4 transition is picking up emission from material at smaller scales, closer to the protostellar source(s), and thus related to the luminosity of the protostellar source(s).
\ce{HNC}, and all three transitions of \ce{^{13}CO} also show relation to $L_{\rm bol}$, as previously found for a larger sample by e.g., \citet{irene2013} and \citet{yildiz2013}.
The other molecules do not present any correlation to bolometric luminosity, not even \ce{SO} and \ce{CH3OH} which are expected to trace shocks, or the higher transitions of \ce{H2CO} 3--2 which trace warmer gas.
The lack of correlation could be due to the low number of detections, and are further explored through statistical analysis in Section~\ref{subsec:stats}.

The peak antenna temperatures compared with $M_{\rm env} / L_{\rm bol}$ are presented in Fig.~\ref{fig:mlpeak}.
There appears to be no relation between any of the observed molecular line peaks and $M_{\rm env} / L_{\rm bol}$.
This indicates that when the amount of mass being illuminated, and thus heated, by the central protostellar system is taken into account, the observed protostellar systems present similar peak antenna temperatures.

These results can be summarized as follows. The bulk of the cold envelope gas is traced by \ce{DCO+} and \ce{H2CO} 3$_{0,3}$--2$_{0,2}$ while the warm gas is traced by \ce{DCN} and the two higher-lying transitions of \ce{H2CO}. \ce{c-C3H2} and \ce{C2H} trace the warm irradiated outflow cavity, rather than envelope material.

\begin{figure*}
	\centering
	\includegraphics[width=0.98\textwidth]{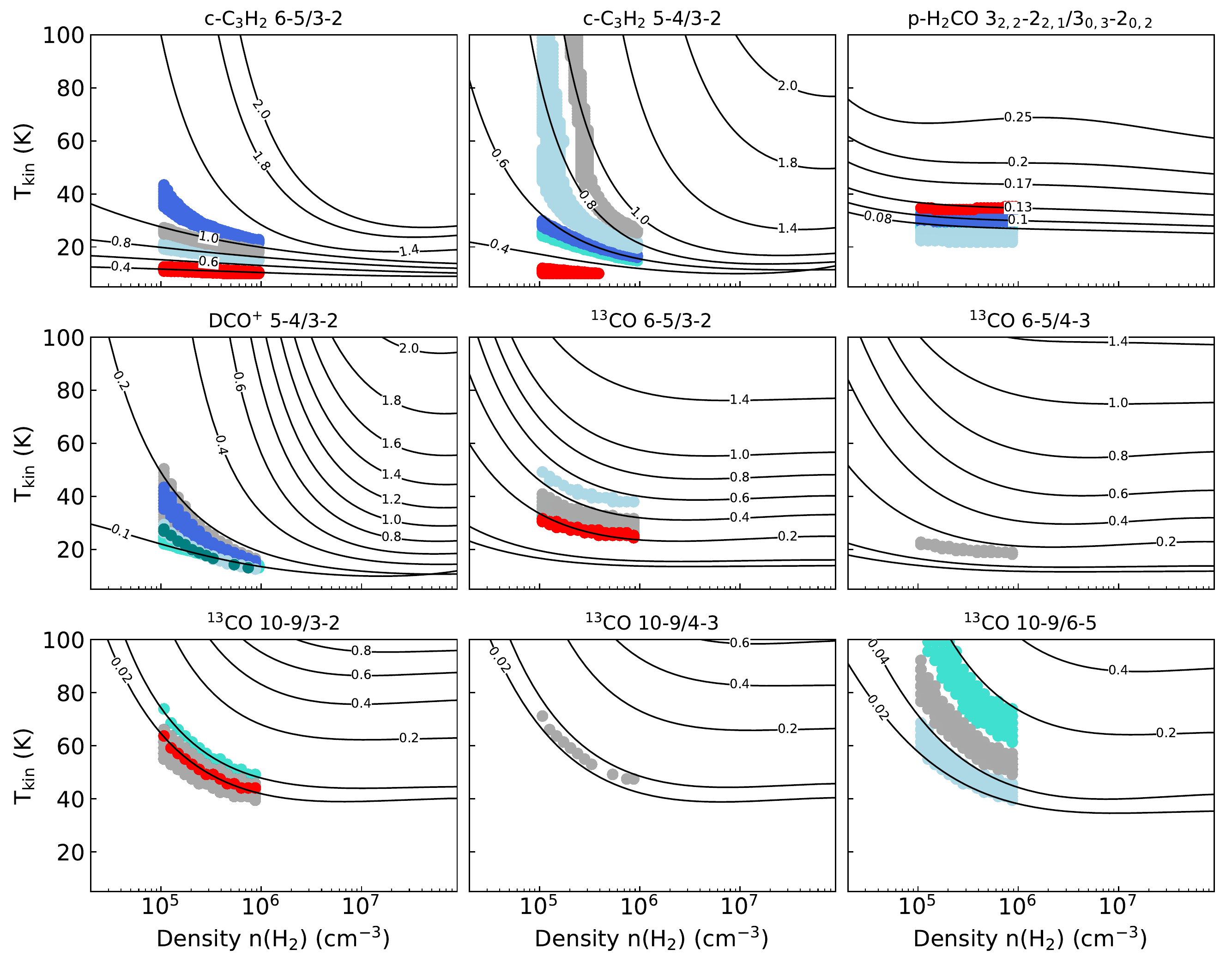}
	\caption{Calculated line ratios for \ce{c-C3H2}, \ce{H2CO}, \ce{DCO+} and \ce{^{13}CO}. The black lines show the modelled ratios assuming a column density of 10$^{12}$ cm$^{-2}$ for \ce{c-C3H2}, \ce{H2CO} and \ce{DCO+}, and 10$^{16}$ cm$^{-2}$ for \ce{^{13}CO}. For the \ce{^{13}CO} 6--5/3--2 and 6--5/4--3 panels, the two lower lines show the 0.02 and 0.04 ratios. The shaded areas show the results for the gas temperature calculations and adopted \ce{H2} density range for individual systems: L1448N (cyan), NGC1333 SVS13 (blue), NGC1333 IRAS7 (light blue), NGC1333 IRAS1 or IRAS 03292+3039 (gray) and NGC1333 IRAS5 Per52 or NGC1333 SK1 (red).}
	\label{fig:ratios}
\end{figure*}

\begin{figure*}
	\centering
	\includegraphics[width=0.98\textwidth]{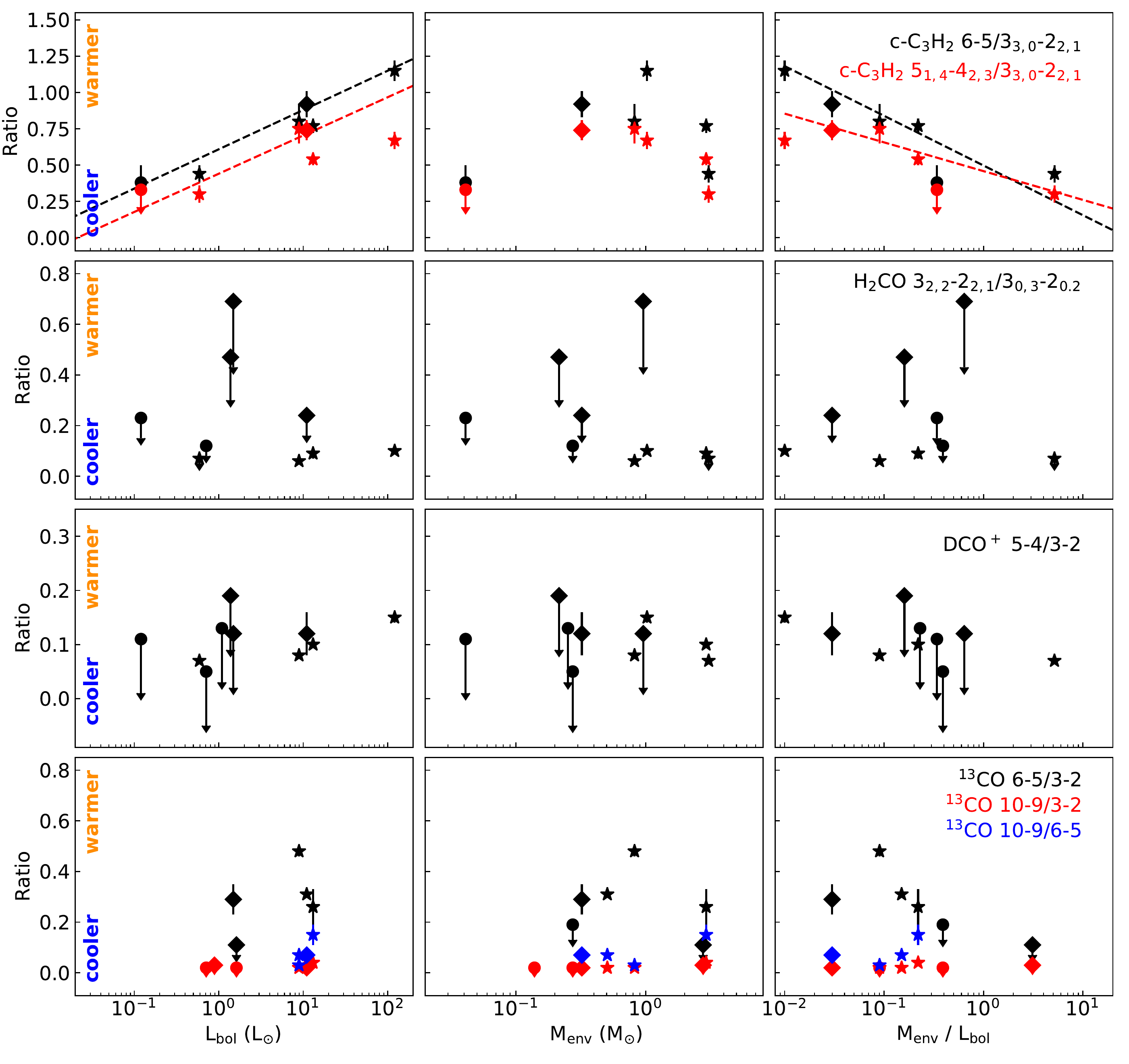}
	\caption{Calculated molecular line ratios of \ce{c-C3H2} (top row), \ce{H2CO} (second row), \ce{DCO+} (third row) and \ce{^{13}CO} (bottom row) compared to the system parameters bolometric luminosity $L_{\rm bol}$ (left column), envelope mass $M_{\rm env}$ (middle column) and the mass to luminosity ratio $M_{\rm env} / L_{\rm bol}$ (right column). The black and red dashed lines in the top row are linear fits to the data for the 6-5/3$_{\rm 3,0}$-2$_{\rm 2,1}$ and 5$_{\rm 1,4}$-4$_{\rm 2,3}$/3$_{\rm 3,0}$-2$_{\rm 2,1}$ ratios, respectively. Circles, diamonds and stars show single, close binary and wide multiple protostellar systems, respectively. The \ce{c-C3H2} ratios appear to be somewhat related to luminosity, whereas the ratios from \ce{H2CO} and \ce{DCO+} are constant regardless of system parameter.}
	\label{fig:param_vs_ratio}
\end{figure*}

\begin{figure*}
	\centering
	\includegraphics[width=\textwidth]{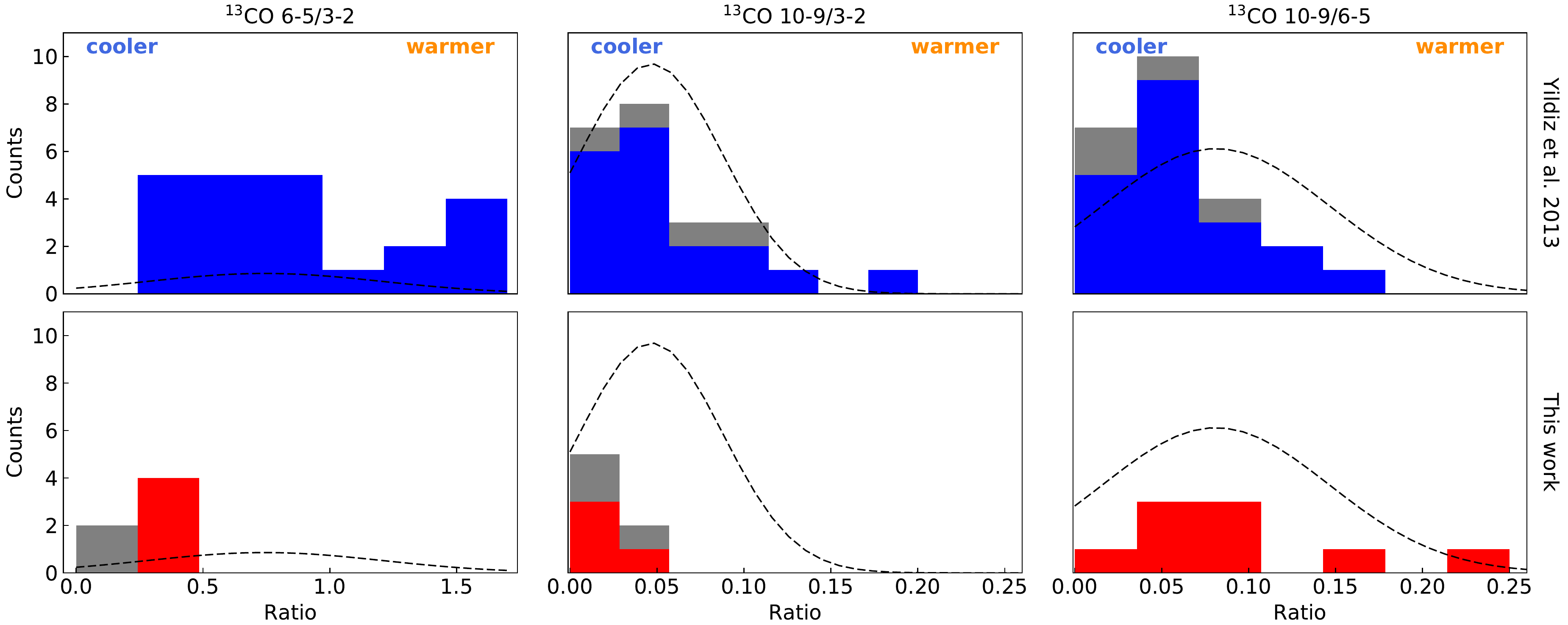}
	\caption{\ce{^{13}CO} ratio stacked histogram including JCMT and \textit{Herschel} data from \citet{yildiz2013} (top row) and this work (bottom row). Upper limits are shown in gray. Ratios from \citet{yildiz2013} are for the systems in the Water In Star-forming regions with \textit{Herschel} (WISH) key project \citep{ewine2011}, which do not overlap with the sample studied in this work. The black dashed line shows a Gaussian distribution fit to the total sample of \ce{^{13}CO} ratios.}
	\label{fig:histo13co}
\end{figure*}

\subsection{Line ratios and implied physical conditions}
\label{subsec:ratios}

The gas temperature structure can be probed with the several transitions of \ce{c-C3H2}, \ce{H2CO} and \ce{DCO+} that were observed.
In addition, the ratio of \ce{DCN}/\ce{DCO+} can be used to compare the warm (\ce{DCN}) and cold gas (\ce{DCO+}) from the envelope. 
Since the two transitions of \ce{DCO+} were observed with a different beam (3--2: 28.7$\arcsec$; 5--4: 18$\arcsec$), a beam dilution correction factor of 0.39 is applied to the 5--4 transition.
Not considering the different beam sizes of the observations, would result in the \ce{DCO+} ratio being overestimated by a factor of 2 to 3.
The line ratios of \ce{c-C3H2}, \ce{H2CO}, \ce{DCO+} and \ce{DCN}/\ce{DCO+} are listed in Table~\ref{tab:ratios}.
While two transitions of \ce{DCN} were observed, ratios can only be obtained for half the systems due to low detection rates, of which three are upper limits.
These ratios are thus also listed for each system where both lines were detected in Table~\ref{tab:ratios} but not discussed further.

High-$J$ CO lines ($J_{\rm u}$ $>$ 3) can be used as diagnostics of temperature and density, as well as UV photon-heated gas (e.g., \citealt{yildiz2012,yildiz2015}).
Hence the \ce{^{13}CO} 4--3 and 6--5 observations presented in this work are compared with peak antenna temperatures from the JCMT and \textit{Herschel} HIFI observations.
For this purpose, the spectra from the JCMT and APEX data were smoothed to the beam of the \textit{Herschel} HIFI observations (19.25$\arcsec$).
Only part of our sample has data from the JCMT and/or \textit{Herschel} HIFI (Table~\ref{tab:13co}).
Thus, only those systems are considered for \ce{^{13}CO} ratios (Table~\ref{tab:13coratios}).

Using RADEX \citep{vandertak2007}, non-LTE excitation and radiative transfer calculations were performed to study the variation of the \ce{c-C3H2}, \ce{H2CO}, \ce{DCO+} and \ce{^{13}CO} ratios with \ce{H2} density and temperature (Fig.~\ref{fig:ratios}).
The molecular data files for the RADEX calculations were obtained from the Leiden Atomic and Molecular Database \citep[LAMDA;][]{schoier2005}. 
The collisional rate coefficients for \ce{DCO+} are based on the results of \citet{botschwina1993} and \citet{flower1999}. For \ce{c-C3H2}, \ce{H2CO} and \ce{^{13}CO}, the collisional rate coefficients are based on \citet{chandra2000}, \citet{wiesenfeld2013} and \citet{yang2010}, respectively.

Ratios are not calculated for the systems with non-detections in both transitions used in the ratio (Tables~\ref{tab:dcodcn}, \ref{tab:c3h2}, \ref{tab:h2co_3-2} and \ref{tab:13co}) and are thus not considered in the following analysis (Table~\ref{tab:ratios} and \ref{tab:13coratios}).
Upper limits are given when one of the transitions is a non-detection (Table~\ref{tab:ratios}), and are not considered for the linear fits but are shown in the figures for reference.

Kinetic gas temperatures are derived from the \ce{c-C3H2}, \ce{H2CO}, \ce{DCO+} and \ce{^{13}CO} ratios by averaging over a range of \ce{H2} densities, and are listed in Table~\ref{tab:Tkin}.
An \ce{H2} density $n_{\ce{H2}}$ range between 10$^{5}$ to 10$^{6}$ cm$^{-3}$, typical in the envelopes of embedded protostellar objects on the scales of the beam, is assumed for the \ce{c-C3H2}, \ce{H2CO}, and \ce{DCO+} calculations. 
A column density of 10$^{12}$ cm$^{-2}$ is adopted for the \ce{DCO+}, \ce{c-C3H2} and \ce{H2CO} line ratio calculations, ensuring that the lines are optically thin.
For \ce{^{13}CO}, a column density of 10$^{16}$ cm$^{-2}$ is adopted for a line width of 2 km~s$^{-1}$, which produces optically thin emission. \citet{yildiz2012} found a column density of 10$^{17}$ cm$^{-2}$ for a line width of 10 km~s$^{-1}$ toward NGC1333 IRAS4A and IRAS4B. Thus our adopted \ce{^{13}CO} column density is reasonable. Adopting a column density of 10$^{17}$ cm$^{-2}$ for a line width of 2 km~s$^{-1}$ would provide optically thick emission.

Figure~\ref{fig:param_vs_ratio} compares the calculated ratios with envelope mass $M_{\rm env}$, bolometric luminosity $L_{\rm bol}$, and $M_{\rm env} / L_{\rm bol}$.
Since molecular line ratios are related to gas temperature (Fig.~\ref{fig:ratios}), the comparison serves to determine relations between temperature and system parameters. 

With respect to $L_{\rm bol}$, both ratios of \ce{c-C3H2} show good correlation with luminosity. 
On the other hand, \ce{c-C3H2} shows an anti-correlation with $M_{\rm env} / L_{\rm bol}$, to be expected based on the correlation with $L_{\rm bol}$.
Both \ce{c-C3H2} ratios point to gas temperatures between 10 and 60 K, with the ratios of NGC1333 IRAS7 and NGC1333 IRAS1 pointing to higher temperatures at $n_{\ce{H2}}$ of a few 10$^{5}$ cm$^{-3}$.
These results suggest that \ce{c-C3H2} is dependent on the protostellar system luminosity. 
This makes sense if it is considered that \ce{c-C3H2} traces the outflow cavity, which is irradiated by the central protostar, and thus the $L_{\rm bol}$. 
A higher luminosity will lead to higher temperatures traced by \ce{c-C3H2}. If the envelope mass is large, however, then more material needs to be heated by the protostar and the gas temperature traced by \ce{c-C3H2} will be lower.
Since the density in the outflow cavity is expected to decrease and the outflow to become hotter as protostars evolve from Class 0 to I (e.g., \citealt{nisini2015,mottram2017}), the results of \ce{c-C3H2} toward our sample may indicate an evolutionary effect.
The separation of the sources in multiple systems or their coevality does not show any effect on the temperature traced by the \ce{c-C3H2} ratio (Fig.~\ref{fig:param_vs_ratio}). 
While there is no difference in temperature between close binaries and wide multiples, the effect of multiplicity, on the other hand, cannot be fully determined, since none of the single protostars in our sample present \ce{c-C3H2} detections.

The ratios of \ce{H2CO}, \ce{DCO+} and \ce{^{13}CO} are quite constant in relation to all system parameters, and suggest overall cooler temperatures.
System type, region and evolutionary stage do not present any correlation either.
The ratios of \ce{H2CO} and \ce{DCO+} indicate temperature between 10 and 60 K.
Considering higher $n_{\ce{H2}}$, alters the temperature by only a few degrees, otherwise the temperatures stay mainly constant.
Thus, all embedded protostars appear to have envelope gas with similar, and relatively cold, temperatures regardless of their multiplicity (Fig.~\ref{fig:param_vs_ratio}).

The ratios obtained from \ce{^{13}CO} vary little among the systems, regardless of system parameters, multiplicity and evolutionary stage. 
The low ratios suggest temperatures typically below 60 K, with only L1448 N showing temperatures closer to 100 K at $n_{\ce{H2}}$ of a few 10$^{5}$ cm$^{-3}$.
This is consistent with the work of \citet{yildiz2015}, which found typical gas temperatures of 30 to 50 K toward protostellar systems.
Given that high-$J$ \ce{^{13}CO} lines trace UV photon-heated gas, the low ratios and derived temperatures would indicate that the envelope is not being heated out to large radii.
Furthermore, the lack of line detection toward some of the systems would suggest that \ce{^{13}CO} emission is concentrated closer to the source of heating, and thus is being diluted in the APEX beam, and even more so when smoothed to the \textit{Herschel} HIFI beam.

In order to determine whether our sample of protostellar systems is particularly cold, we construct a histogram of \ce{^{13}CO} ratios from Appendix C of \citet{yildiz2013} and those derived from the observations in this work (Fig.~\ref{fig:histo13co}).
In total the sample includes 33 protostellar systems, 26 from the work of \citet{yildiz2013}, and 7 from this work.
The sample from \citet{yildiz2013} includes protostellar systems from different star forming regions, but does not overlap with the sample in this work.
The histogram is then fit with a Gaussian distribution to find the mean ratio and standard deviation.
For the \ce{^{13}CO} 6--5/3--2 ratio, the mean ratio is 0.74 with a standard deviation of 0.46, and the ratios indicating a range of temperatures.
For 10--9/3--2 and 10--9/6--5, the ratios are below 0.3 and indicate cool temperatures in general.
For the \ce{^{13}CO} 10--9/3--2 ratio, the mean ratio is 0.05 with a standard deviation of 0.04, whereas for 10--9/6--5, the mean ratio and standard deviation are 0.08 and 0.07, respectively.
The upper limits would decrease the mean ratio for the \ce{^{13}CO} 10--9/3--2 and 10--9/6--5 ratios, but not so much for the 6--5/3--2 ratio.

In summary, the cool envelope temperatures found in this work are not from particularly cold protostellar systems, nor is the Perseus molecular cloud producing uncommon protostars.
It would seem, instead, that the central protostar does not extensively heat the envelope to high temperatures, and that single protostars do not heat the envelope differently than multiple protostellar systems.

\begin{table*}
	\centering
	\caption{Peak main beam temperature line ratios}
	\begin{tabular}{ccccccc}
		\hline \hline
		System & \ce{c-C3H2} & \ce{c-C3H2} & \ce{H2CO} & \ce{DCO+}  &  \ce{DCN}  & \ce{DCN}/\ce{DCO+} \\
		& 6--5 / 3$_{3,0}$--2$_{2,1}$ & 5$_{1,4}$--4$_{2,3}$ / 3$_{3,0}$--2$_{2,1}$ & 3$_{2,2}$--2$_{2,1}$ / 3$_{0,3}$--2$_{0,2}$ & 5--4/3--2 & 5--4/3--2 & 3--2 \\
		\hline
		\multicolumn{7}{c}{Wide multiples} \\
		\hline
		L1448N  & 0.77 $\pm$ 0.05 & 0.54 $\pm$ 0.04 & 0.09 $\pm$ 0.02 & 0.1 $\pm$ 0.01 & 0.26 $\pm$ 0.09 & 0.08 $\pm$ 0.01 \\
		SVS13  & 1.15 $\pm$ 0.07 & 0.67 $\pm$ 0.06 & 0.1 $\pm$ 0.01 & 0.15 $\pm$ 0.03 & 0.35 $\pm$ 0.10 & 0.24 $\pm$ 0.02 \\
		Per63 &  …  &  …  &  <0.47  & $<$ 0.19  &  … &  <0.36  \\
		Per52 & 0.38 $\pm$ 0.12 &  <0.33  &  <0.23  & $<$ 0.11  &  … &  <0.21  \\
		IRAS7  & 0.8 $\pm$ 0.12 & 0.75 $\pm$ 0.1 & 0.06 $\pm$ 0.01 & 0.08 $\pm$ 0.01 & $<$ 0.41 & 0.12 $\pm$ 0.01 \\
		B1-b  & 0.44 $\pm$ 0.06 & 0.3 $\pm$ 0.06 &  <0.07  & 0.07 $\pm$ 0.01 & $<$ 0.35 & 0.1 $\pm$ 0.01 \\
		\hline
		\multicolumn{7}{c}{Close binaries} \\
		\hline
		IRAS1  & 0.92 $\pm$ 0.09 & 0.74 $\pm$ 0.07 &  <0.24  & 0.12 $\pm$ 0.1 &  … &  <0.15  \\
		IRAS 03282  &  …  &  …  &  <0.69  & $<$ 0.12  & $<$ 0.76 & 0.1 $\pm$ 0.03 \\
		\hline
		\multicolumn{7}{c}{Singles} \\
		\hline
		IRAS 03271  &  …  &  …  &  …  & …  &  … &  …  \\
		Per25  &  …  &  …  &  …  & $<$ 0.13  &  … &  <0.52  \\
		SK1  &  …  &  …  &  <0.12  & $<$ 0.05  &  … &  <0.1  \\
		\hline
	\end{tabular}
	\label{tab:ratios}
\end{table*}

\begin{table*}
	\centering
	\caption{\ce{^{13}CO} peak main beam temperature line ratios}
	\begin{tabular}{cccccc}
		\hline
		\hline
		System &  6--5/3--2 & 6--5/4--3 &  10--9/3--2 & 10--9/4--3  &  10--9/6--5  \\
		\hline
		\multicolumn{6}{c}{Wide multiples} \\
		\hline
		L1448N & 0.26 $\pm$ 0.07 & & 0.04 $\pm$ 0.004 & & 0.15 $\pm$ 0.04 \\
		IRAS7\tablefootmark{a} & 0.48 $\pm$ 0.02 & & 0.02 $\pm$ 0.004 & & 0.03 $\pm$ 0.01 \\
		PER8+PER55 & 0.31 $\pm$ 0.01 & & 0.02 $\pm$ 0.003 & & 0.07 $\pm$ 0.01 \\
		\hline
		\multicolumn{6}{c}{Close binaries} \\
		\hline
		IRAS1 & 0.29 $\pm$ 0.06 & & 0.02 $\pm$ 0.01 & & 0.07 $\pm$ 0.02 \\
		IRAS 03282 &  & ... & & ... & ... \\
		IRAS 03292 & & $<$0.11 & & $<$0.03 & ... \\ 
		\hline
		\multicolumn{6}{c}{Singles} \\
		\hline
		IRAS 03271 & ... & & $<$0.02 & & ... \\
		Per25 & & & & & ... \\
		SK1 & $<$0.19 & & $<$0.02 & & ...\\ 
		\hline
	\end{tabular}
\tablefoot{\tablefoottext{a}{Includes only Per18 and Per21.}}
\label{tab:13coratios}
\end{table*}

\begin{table*}
	\centering
	\caption{Derived $T_{\rm kin}$ from line ratios averaged over \ce{H2} density range}
	\begin{tabular}{c c c c c c c c}
		\hline \hline
		System & \ce{c-C3H2}\tablefootmark{a} & \ce{c-C3H2}\tablefootmark{a} & \ce{H2CO}\tablefootmark{a} & \ce{DCO+}\tablefootmark{a} & \ce{^{13}CO}\tablefootmark{a} & \ce{^{13}CO}\tablefootmark{a} & \ce{^{13}CO}\tablefootmark{a} \\
		& 6--5 / 3$_{3,0}$--2$_{2,1}$ & 5$_{1,4}$--4$_{2,3}$ / 3$_{3,0}$--2$_{2,1}$ & 3$_{2,2}$--2$_{2,1}$ / 3$_{0,3}$--2$_{0,2}$ & 5--4 / 3--2 & 6--5/3--2 & 10--9/3--2 & 10--9/6--5 \\
		& $T_{\rm kin}$ (K) & $T_{\rm kin}$ (K) & $T_{\rm kin}$ (K) & $T_{\rm kin}$ (K) & $T_{\rm kin}$ (K) & $T_{\rm kin}$ (K) & $T_{\rm kin}$ (K) \\
		\hline
		\multicolumn{8}{c}{Wide multiples} \\
		\hline 
		L1448N & 18 $\pm$ 2 & 20 $\pm$ 3 & 30 $\pm$ 2 & 19 $\pm$ 3 & 30 $\pm$ 3 & 58 $\pm$ 7 & 86 $\pm$ 16 \\
		SVS13 & 30 $\pm$ 6 & 22 $\pm$ 4 & 31 $\pm$ 1 & 26 $\pm$ 8 & ... & ... & ... \\
		Per63 & … & … & $<$180 & $<$29 & ... & ... & ... \\
		Per52 & 11 $\pm$ 1 & $<$10 & $<$61 & $<$21 & ... & ... & ... \\
		IRAS7 & 18 $\pm$ 2 & 54 $\pm$ 30 & 24 $\pm$ 1 & 17 $\pm$ 3 & 42 $\pm$ 3 & 51 $\pm$ 6 & 51 $\pm$ 7 \\
		B1-b & 11 $\pm$ 1 & 10 $\pm$ 7 & $<$25 & 21 $\pm$ 5 & ... & ... & ... \\
		Per8+Per55 & ... & ... & ... & ... & 33 $\pm$ 2 & 50 $\pm$ 6 & 66 $\pm$ 10 \\
		\hline
		\multicolumn{8}{c}{Close binaries} \\
		\hline
		IRAS1 & 22 $\pm$ 3 & 81 $\pm$ 49 & $<$64 & 25 $\pm$ 8 & 33 $\pm$ 3 & 50 $\pm$ 6 & 65 $\pm$ 10 \\
		IRAS 03282 & … & … & $<$180 & $<$26 & ... & ... & ... \\
		IRAS 03292 & ... & ... & ... & ... & $<$20\tablefootmark{b} & $<$57\tablefootmark{b} & ... \\
		\hline
		\multicolumn{8}{c}{Singles} \\
		\hline
		IRAS 03271 & … & … & … & ... & ... & $<$51 & ... \\
		Per25 & … & … & … & $<$23 & ... & ... & ... \\
		SK1 & … & … & $<$35 & $<$13 & $<$27 & $<$51 & ... \\
		\hline
	\end{tabular}
	\tablefoot{\tablefoottext{a}{Assuming a \ce{H2} density range of 10$^{5}$ to 10$^{6}$ cm$^{-3}$}
		       \tablefoottext{b}{Kinetic temperatures are for the 6--5/4--3 and 10--9/4--3 ratios.}}
	\label{tab:Tkin}
\end{table*}

\subsection{Statistical analysis}
\label{subsec:stats}
In order to determine quantitatively if there is a relation between the observed line peaks and derived quantities, and the system parameters, the Generalized Kendall's rank correlation is used \citep{isobe1986}.
This method measures the degree of association between two quantities which contain upper limits (censored data), with the null hypothesis being that the values are uncorrelated.
Thus, if the significance level $p >$ 0.05, the probability of the values being correlated is less than 3$\sigma$, while $p <$ 0.05 indicates a correlation at better than 3$\sigma$ significance.
The standard normal score $z$ and the significance level $p$ of the Generalized Kendall's rank correlation are listed in Table~\ref{tab:stats}, with values indicating a correlation highlighted in bold text.
The significance level is calculated from the standard normal score $z$ by the relation
\begin{equation}
p = 1 - 0.5*(1 + erf\left(\frac{|z|}{\sqrt{2}}\right)),
\end{equation}
where $erf(x)$ is the error function.

The results confirm the correlations seen by eye (Sec.~\ref{subsec:peakparam}).
The peak antenna temperatures of the molecules tracing cold gas, namely \ce{DCO+} and \ce{H2CO} are associated with system envelope mass, but not with luminosity.
On the other hand, the peak antenna temperatures of \ce{c-C3H2} and \ce{C2H}, both tracers of warm gas, are correlated with luminosity but not system envelope mass. 
The three transitions of \ce{^{13}CO} present correlation with luminosity, while the 6--5 transition also presents a correlation with the mass to luminosity ratio.
The 6--5 / 3$_{3,0}$--2$_{2,1}$ ratio of \ce{c-C3H2} shows correlations with luminosity and the mass to luminosity ratio, but not envelope mass. 
The \ce{DCN}/\ce{DCO+} and \ce{^{13}CO} 10--9/3--2 ratios present correlation with envelope mass, but not the other system parameters.

\section{Discussion} 
\label{sec:diss}

\begin{figure}
	\centering
	\includegraphics[width=0.98\columnwidth]{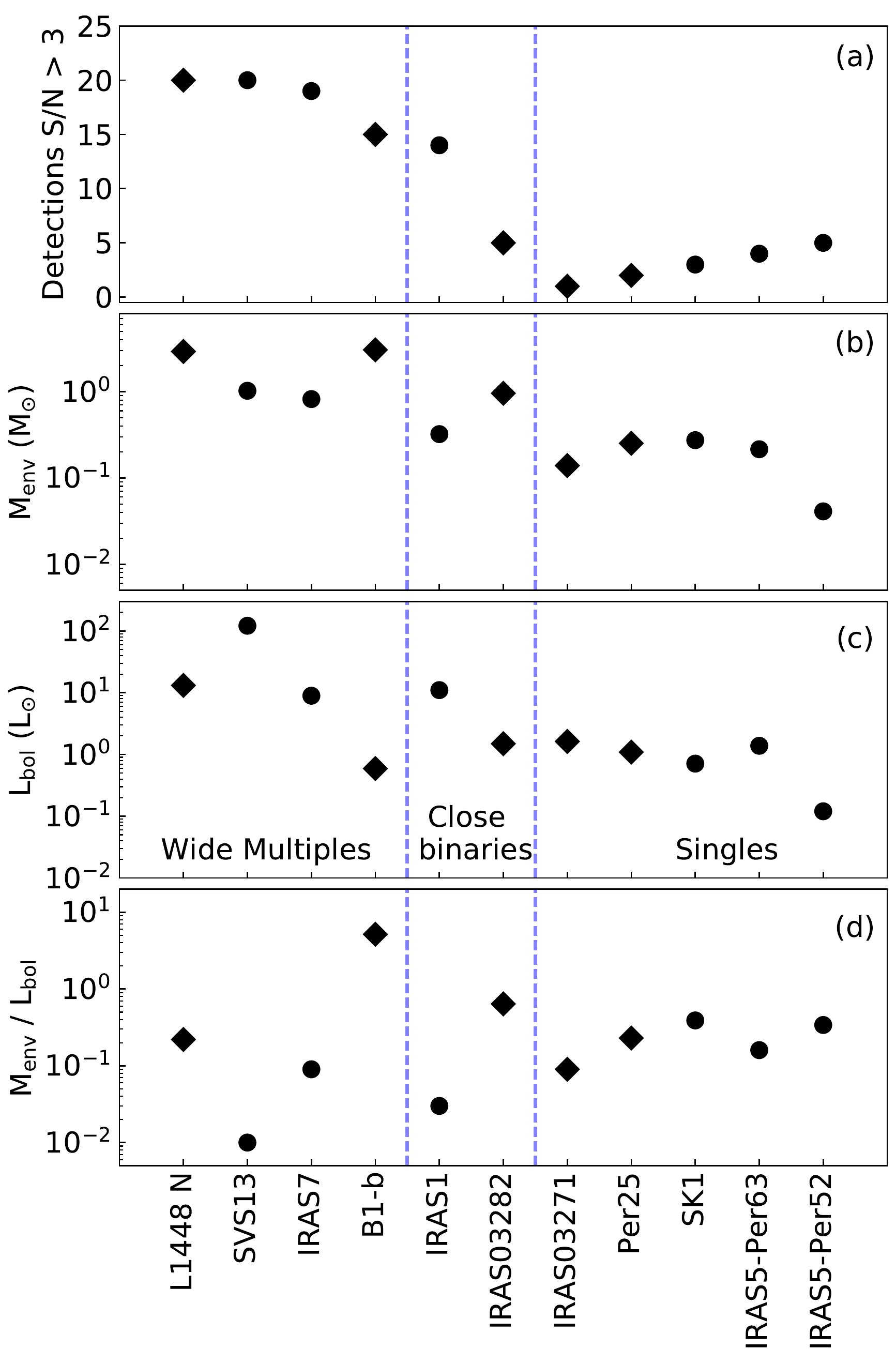}
	\caption{Number of line detections (a), envelope mass $M_{\rm env}$ (b), bolometric luminosity $L_{\rm bol}$ (c), and envelope mass to luminosity ratio $M_{\rm env}$ / $L_{\rm bol}$ (d) of each system in our sample compared to the region, clustered (circles) or non-clustered (diamonds), and system type. NGC1333 IRAS5 is considered here as two single protostellar systems, Per63 and Per52. }
	\label{fig:sourcevschem}
\end{figure}

\subsection{Observed line detections}

The observations appear to show a tendency for wide multiple protostellar systems (separations $>$7$\arcsec$) to have more molecular line detections with strong peak intensities than close binaries (Fig.~\ref{fig:sourcevschem}). 
In contrast, single protostars present very weak molecular line emission.
There seems to be a relation, however, between the envelope mass and the number of molecular line detections.
As noted in Sect.~\ref{subsec:peakparam}, the ratio $M_{\rm env} / L_{\rm bol}$ correlates with evolutionary stage for the close binary and single systems. However, the ratio, and thus the evolutionary stage, does not seem to have relation to the number of line detections (Fig.~\ref{fig:sourcevschem}d).
These relations appear to not be affected by clustering either.
However, the envelope mass is larger for wide multiple protostellar systems in contrast to that of close binary and single protostars.
Core mass would then be more related to formation of non-coeval wide multiple protostellar systems.
There is no apparent relation between the bolometric luminosity $L_{\rm bol}$ and the number of line detections (Fig.~\ref{fig:sourcevschem}c) nor their strength (Fig.~\ref{fig:Lbolpeak}).
The systems L1448 N, NGC1333 IRAS7 and NGC1333 SVS13, which have combined bolometric luminosities above 8 L$_{\odot}$, show strong detections of all the molecular species.
Clustering does not seem to particularly enhance the line strength or number of line detections in the envelope (Fig.~\ref{fig:sourcevschem}). 
L1448 N and B1-b are both located in non-clustered environments and present the same chemical richness as the multiple protostars located in NGC1333, which is a clustered region.
Both Class 0 and I systems are present in wide multiples, close binaries and single protostellar systems, but no effect is seen on the line detections. 
It must be highlighted, however, that the sample size presented in this work is small and more protostellar systems would be needed to determine if the trends found in this work are real or product of the small sample size. 

\subsection{Envelope gas temperature and multiplicity} 
The \ce{H2CO} and \ce{DCO+} ratios point to cool envelopes ($T_{\rm gas}<$ 60 K at scales of 7000 AU) for all the systems in our sample, regardless of multiplicity, clustering and evolutionary stage.
Only \ce{c-C3H2} presents a correlation with bolometric luminosity and an anti-correlation with $M_{\rm env} / L_{\rm bol}$. Since a formation path of \ce{c-C3H2} is through the breakdown of large hydrocarbons by UV photons, these correlations could be related to the amount of material that is irradiated by the central protostar, with more luminous protostars irradiating more material. Alternatively, it could be an evolutionary effect where the jet becomes hotter while the envelope and outflow cavity mass decreases, resulting in more material being irradiated by UV photons (e.g., \citealt{nisini2015,mottram2017}).
The low \ce{^{13}CO} ratios, and derived temperature ($T_{\rm gas}<$ 100 K at scales of $\sim$4500 AU), suggest that the envelope and outflow cavity are not being UV photon-heated out to large extents.

The results of \citet{maret2004} and \citet{koumpia2016} support the possibility that the envelope gas is not being heated out to large extents. \citet{maret2004} observed L1448 N, NGC1333 IRAS4A and IRAS4B using the single-dish IRAM 30-meter telescope and the JCMT, tracing scales of 2500 to 4000 AU (HPBW = 11$\arcsec$ to 17$\arcsec$), while \citet{koumpia2016} observed NGC1333 IRAS4 with the JCMT (HPBW = 15$\arcsec$, $\sim$3500 AU). Using LVG modelling under non-LTE conditions, \citet{maret2004} derive temperatures of 50 K for NGC1333 IRAS4A, 80 K for NGC1333 IRAS4B and 90 K for L1448 N. The results presented here for L1448 N find gas temperature of 30 K at 7000 AU scales and 65 K at 4500 AU scales (Table~\ref{tab:Tkin}). The peak gas temperatures found by \citet{koumpia2016} for NGC1333 IRAS4 are in agreement with the results of \citet{maret2004}. However the gas temperature distribution from \citet{koumpia2016} shows that the gas temperature drops to 40 K at $\sim$2000 AU away from NGC1333 IRAS4B, and that the 60 K gas is located along the outflow cavity of NGC1333 IRAS4A, consistent with the findings of \citet{yildiz2012}, with the gas temperature droping to 40 K at a distance of $\sim$5000 AU.
These results thus indicate that while the embedded protostar can heat up the surrounding gas to temperatures of 50 to 100 K, it can only do so out to about 1000 AU or along low density structures such as the outflow cavity. In contrast wide multiple protostellar systems are observed to have much larger separations (Table~\ref{tab:source}), with sources not likely to be located within the outflow cavity.

Non-detection within the APEX beam of molecules such as \ce{c-C3H2} and \ce{H2CO}, which trace the outflow cavity and envelope gas, respectively, would point to cooler envelopes and outflow cavities at scales of 7000 AU.
In addition, \ce{DCO+}, a cold gas tracer, is detected toward all but one system (IRAS 03271+3013, single).
Non-detections of warm gas tracers in addition to the presence of \ce{DCO+} toward single and close binary protostellar systems could be explained by these systems being somewhat cooler than wide multiple protostellar systems. 
Another possibility could be the evolutionary dispersal of envelope material which would cause the warm gas to be located closer to the protostellar source(s) and thus diluted in the observing beam.
Interestingly enough, \ce{H2CO} 3$_{0,3}$--2$_{0,2}$ and \ce{DCO+} appear to increase with an increase in envelope mass. In other words, \ce{H2CO} 3$_{0,3}$--2$_{0,2}$ and \ce{DCO+} present stronger emission in wide multiple protostars. This could indicate the presence of massive reservoirs of cold gas in these systems.

External heating does not seem to play a significant role, as evidenced by a lack of difference in all aspects between systems in clustered and non-clustered regions.
However it is not clear why close binaries present less molecular line emission than wide multiple protostars, given that some sources of the latter are often close binaries themselves (Fig.~\ref{fig:sample}).
A possible explanation would be the sensitivity and beam filling factor of the observations. However, this would not explain why NGC1333 IRAS1 has more line detections than NGC1333 IRAS5 Per63 when both have sensitivities of $\sim$30 mK.
The results presented in this work indicate that there is no significant difference in envelope temperature between single, close binary and wide multiple protostellar systems.
The only difference temperature-wise arises from \ce{c-C3H2}, which traces the outflow cavity.
This is consistent with previous studies that find the heating by protostellar source(s) is mainly channeled through the outflow cavity \citep{yildiz2012,yildiz2015}.

The envelope gas temperatures measured in our sample coincide well with the temperature expectations from \citet{krumholz2006}, 30 K at a few thousand AU, while the results from \citet{koumpia2016} and \citet{maret2004} are consistent with the inner regions closer to the source being heated to 100 K.
However, no extensive heating of the envelope out to thousand AU scales \citep{bate2012} is found in our sample, with gas temperatures above 30 K occurring only in the outflow cavities.
Based on the work presented here, the prediction that heating of the envelope should hinder fragmentation is not consistent with the results and analysis of the sample.
Wide multiple systems tend to be non-coeval and are thus produced by continued fragmentation of the cloud core \citep{murillo2016,sadavoy2017}, but they do not show colder envelope gas temperatures with respect to coeval binary and single protostellar systems.
Thus, if the envelope gas temperature is similar for all surveyed systems, then it is not clear why only some cores have continued fragmentation.
Cloud core mass could provide a possible explanation, since wide multiple systems present higher envelope masses than coeval binary and single protostellar systems, however this would require further studies.

An interesting case is that of L1448 N, which has been found to have undergone recent fragmentation in the disk of one source \citep{tobin2016N}, that is fragmentation at 10 to 100 AU scales.
Although the recent fragmentation could have heated the envelope of this system as suggested by simulations \citep{whitehouse2006,boss2000}, increasing the envelope temperature, it does not seem to have generated a considerable effect on thousand AU scales. Furthermore, the disk is certainly heated by the existing protostars but fragmentation of the disk was not hindered.

\subsection{Accretion bursts}
Accretion of material onto the protostar is not a continuous process, but has been found from observations to be episodic (e.g., \citealt{visser2015,safron2015,frimann2017,hsieh2018}). An accretion burst causes an increase in luminosity, and consequently heating of the envelope gas. The chemical composition of the envelope gas is altered, since cold chemistry molecules (e.g. \ce{DCO+}) move further out and molecules such as \ce{CO} are evaporated off the grains.
While the luminosity of the central protostar will decrease after the accretion burst has passed, it takes about 10$^{4}$ years for the gas to refreeze onto the dust grains \citep{johnstone2013,jorgensen2015}.

\citet{frimann2017} studied the presence of accretion bursts toward embedded protostars in Perseus.
Their sample includes four of the systems studied in this work (L1448 N, NGC1333 SVS13, NGC1333 IRAS1 and IRAS 03282), as well as two sources of NGC1333 IRAS7, Per18 and Per21.
Half of these systems (Per21, IRAS1 and IRAS 03282) present indications of past accretion burst activity, while the others (L1448 N, SVS13 and Per18) do not.
The results presented here do not show any difference among the systems that have evidence of having undergone an accretion burst, and those that have not.
NGC1333 IRAS1 and IRAS 03282, both close binary protostars, are suggested to have undergone an accretion burst. The difference found in this work is mainly that IRAS1 presents emission from warm molecules, while IRAS 03282 only shows emission in cold molecules.
For L1448N and SVS13, both wide multiple protostars, there is also no significant difference, since both systems present the same number of line detections, with similar peak intensities except for \ce{DCO+} which is stronger in L1448N.
Analyzing NGC1333 IRAS7 Per18 and Per21 (Appendix~\ref{app:spobs}) separately also points to a lack of significant difference among both sources. However, it must be noted that given their separation of 13$\arcsec$, the spectra of each source is contaminated by emission from the other.

\section{Conclusions}
\label{sec:conc}
Single-dish observations of a sample of 12 embedded protostellar systems in Perseus are presented here.
The observations targeted molecular line emission that trace the cold, warm and UV-heated gas of the observed protostellar systems.
The sample included wide multiple protostellar systems (separation $\geq$ 7$\arcsec$, or $\geq$1600 AU), close binary protostars (separation$<$2$\arcsec$, or $<$470 AU) and single protostars, located in clustered and non-clustered environments, spanning Class 0 and I objects, and containing coeval and non-coeval systems.
The results presented in this work examine the relationship between fragmentation and temperature, since heated gas is expected to suppress fragmentation based on simulations including radiative feedback.

Although the sample presented here is small and there are several upper limits, the envelope gas temperature is found to be similar among multiple and single protostars, regardless of evolutionary stage, coevality or clustering. 
These results suggest that gas temperature may not have as strong a role in suppressing fragmentation as expected from models, a result that was also stated in \citet{offner2010}, who found that temperature does not suppress turbulent fragmentation.
Instead, wide multiple protostellar systems present larger envelope masses and massive cold gas reservoirs in comparison to close binary and single protostars.
It seems, then, that mass, along with other factors such as turbulence, density profile and magnetic fields, rather than envelope gas temperature plays a fundamental role in fragmentation.
Larger, more massive cores could then lead to further fragmentation that forms non-coeval wide multiple systems.

Further interferometric observations of the sample used in this work using the same molecular lines treated here would lend insight into the spatial distribution of the cold, warm and UV heated gas. The gas density and temperature distribution can then be compared to the multiplicity of the protostellar system. 
An additional topic of further research would then be to determine what causes some cores to become more massive than others.

\begin{acknowledgements}
	Astrochemistry in Leiden is supported by the European Union A-ERC grant 291141 CHEMPLAN, by the Netherlands Research School for Astronomy (NOVA), by a Royal Netherlands Academy of Arts and Sciences (KNAW) professor prize.
	JCM acknowledges support from the European Research Council under the European Community’s Horizon 2020 framework program (2014-2020) via the ERC Consolidator grant `From Cloud to Star Formation (CSF)’ (project number 648505).
	AK acknowledges support from the Polish National Science Center grant nr 2016/21/D/ST9/01098.
	We are grateful to the APEX staff for support with these observations. Observing time for the APEX data was obtained via Max Planck Institute for Radio Astronomy, Onsala Space Observatory and European Southern Observatory.
\end{acknowledgements}

\bibliographystyle{aa}
\bibliography{PerseusT.bib}

\begin{appendix}

\section{Single-pointing observations: NGC1333 IRAS7}
\label{app:spobs}

APEX single pointing observations with the heterodyne instruments APEX-1 and APEX-2 were made toward 5 wide (separation $>$7$\arcsec$) multiple protostellar systems in Perseus.
These systems are referred to as wide multiple protostars since the sources span separations grater than 7$\arcsec$ (which can be resolved with \textit{Herschel Space Observatory} PACS photometric maps, \citealt{murillo2016})
One pointing per source in a wide multiple system was observed.
In the main text are the parameters of the lines obtained by averaging the spectra of the individual sources for the corresponding system.
This is because in all cases except NGC1333 IRAS5, the beam of APEX observations partially overlaps another source in the system.

\begin{figure*}
	\centering
	\includegraphics[width=0.95\textwidth]{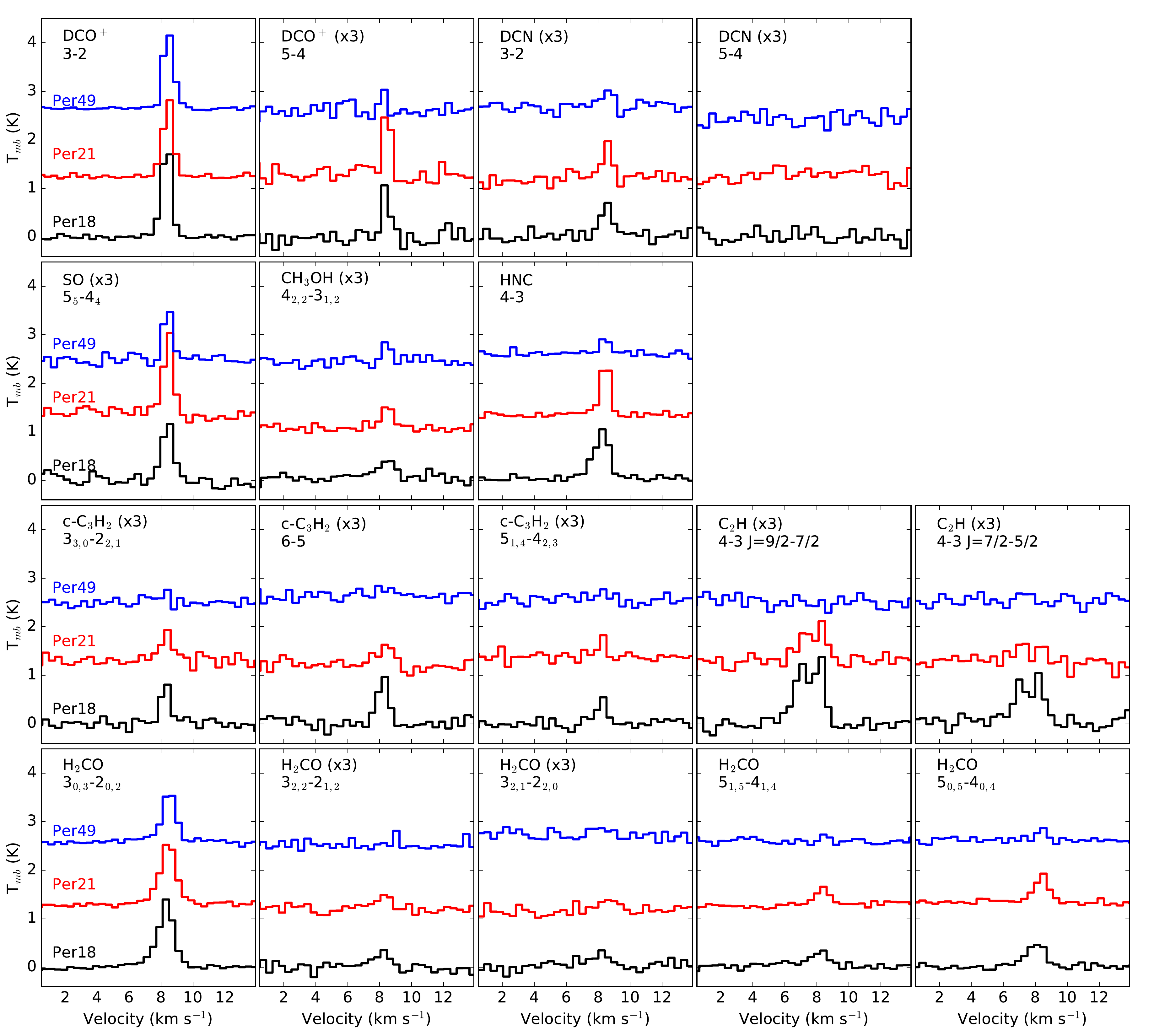}
	\caption{Spectra for the individual sources of NGC1333 IRAS7. Note that some spectra are multiplied by a factor of 3 in order to enhance the line emission features. The spectra of Per21 and Per49 are offset by 1.3 and 2.6 K from that of Per18 for clarity.}
	\label{fig:IRAS7}
\end{figure*}

Here the spectra for the individual sources of the system NGC1333 IRAS7 are presented (Fig.~\ref{fig:IRAS7}).
This provides an example of the similarity of the spectra of the individual sources. 
The spectra of Per21, which is expected to have undergone a recent accretion burst, and Per18, which has no evidence of episodic accretion, can also be compared.
It is also interesting to note that Per49 presents much weaker emission than Per18 and Per21.
The evolutionary stage, multiplicity and bolometric luminosity for each source in the system are listed in Table~\ref{tab:IRAS7param} for reference.

The peak brightness temperature ratios for \ce{c-C3H2}, \ce{H2CO}, \ce{DCO+} and \ce{DCN} are calculated for each source (Table~\ref{tab:IRAS7ratios}). A beam dilution factor of 0.39 is applied to the 5--4 transitions of \ce{DCO+} and \ce{DCN}, since the beam of the observations of the 5--4 transition is smaller than the beam for the 3--2 transition (see Sect.~\ref{sec:obs}). Since the beams of the APEX-1 observations overlap for Per18 and Per21, we note that the ratios for these two sources may be contaminated with emission from each other. Table~\ref{tab:IRAS7ratios} also lists the line ratios for the whole system for the purpose of comparison.

The line ratios for each source in NGC1333 IRAS7 are similar to those derived for the whole system (Table~\ref{tab:IRAS7ratios}), and are compared with the ratio models shown in Fig.~\ref{fig:ratios}. The \ce{DCN} ratios are upper limits due to the non-detection of the 5--4 transition, and instead three times the noise level is used for the ratio. The \ce{DCN} to \ce{DCO+} 3--2 ratio is used to look at the ratio of warm to cold gas in the envelope, with all the ratios well below unity, indicating a larger amount of cold gas.

Gas temperatures were derived from the line ratios assuming an \ce{H2} density range of 10$^{5}$ to 10$^{6}$ cm$^{-3}$ for \ce{c-C3H2}, \ce{H2CO} and \ce{DCO+}. The average derived gas temperatures are listed in Table~\ref{tab:IRAS7temp}. For the envelope gas, \ce{H2CO} indicates temperatures of around 30 K, while \ce{DCO+} ratios point to gas temperatures of about 21 K. For \ce{c-C3H2}, the line ratios indicate higher temperatures, between 20 and 50 K, which is to be expected for material along the outflow cavity. The gas temperatures for each source in the NGC1333 IRAS7 system are similar to the system as a whole (Table~\ref{tab:IRAS7temp}).

The derived temperatures do not show relation with bolometric luminosity or envelope mass. The overall trend is consistent with that of the other systems studied, namely that the envelope gas temperature is similar for all systems, regardless of multiplicity. The source Per21 is thought to have undergone a recent accretion burst \citep{frimann2017}, however no significant difference in the derived ratios and gas temperatures is found. It must be stressed, however, that the values toward Per21 are very likely contaminated by Per18, and viceversa, due to the beam of the APEX-1 observations.

\begin{table*}
	\centering
	\caption{Source parameters for NGC1333 IRAS7}
	\begin{tabular}{cccccccc}
		\hline \hline
		 & Class & Multiplicity & Accretion burst? & 850 $\mu$m peak (Jy~beam$^{-1}$) & $L_{\rm bol}$ (L$_{\odot}$) & $M_{\rm env}$ (M$_{\odot}$) & $M_{\odot}~/~L_{\rm bol}$ (M$_{\odot}$ / L$_{\odot}$) \\
		 \hline
		 Per18 & 0 & Binary & No & 1.76 & 4.8 & 1.0 & 0.21 \\
		 Per21 & 0 & Single & Possible & 1.76 & 3.5 & 1.1 & 0.33 \\
		 Per49 & I & Binary &  & 1.75 & 0.7 & 2.0 & 2.50 \\
		 \hline
	\end{tabular}
	\label{tab:IRAS7param}
\end{table*}

\begin{table*}
	\centering
	\caption{Peak main beam temperature line ratios for NGC1333 IRAS7}
	\begin{tabular}{ccccccc}
		\hline \hline
		 & \ce{c-C3H2} & \ce{c-C3H2} & \ce{H2CO} & \ce{DCO+}  &  \ce{DCN}  & \ce{DCN}/\ce{DCO+} \\
		& 6--5 / 3$_{3,0}$--2$_{2,1}$ & 5$_{1,4}$--4$_{2,3}$ / 3$_{3,0}$--2$_{2,1}$ & 3$_{2,2}$--2$_{2,1}$ / 3$_{0,3}$--2$_{0,2}$ & 5--4/3--2 & 5--4/3--2 & 3--2 \\
		\hline
		Per18 & 1.07 $\pm$ 0.15 & 0.63 $\pm$ 0.10 & 0.08 $\pm$ 0.02 & 0.08 $\pm$ 0.01 & < 0.24 & 0.12 $\pm$ 0.02 \\
		Per21 & 0.65 $\pm$ 0.19 & 0.90 $\pm$ 0.22 & 0.08 $\pm$ 0.03 & 0.16 $\pm$ 0.01 & < 0.18 & 0.17 $\pm$ 0.02 \\
		Per49 & 1.10 $\pm$ 0.45 & < 0.90 & 0.10 $\pm$ 0.03 & 0.05 $\pm$ 0.01 & < 0.50 & 0.08 $\pm$ 0.02 \\
		\hline
		IRAS7\tablefootmark{a} & 0.8 $\pm$ 0.12 & 0.75 $\pm$ 0.1 & 0.06 $\pm$ 0.01 & 0.08 $\pm$ 0.01 & $<$ 0.41 & 0.12 $\pm$ 0.01 \\
		\hline
	\end{tabular}
	\tablefoot{\tablefoottext{a}{Values for the spectra of all three sources averaged together.}}
	\label{tab:IRAS7ratios}
\end{table*}

\begin{table*}
	\centering
	\caption{Derived $T_{\rm kin}$ from line ratios averaged over \ce{H2} for NGC1333 IRAS7}
	\begin{tabular}{ccccc}
		\hline \hline
		 & \ce{c-C3H2} & \ce{c-C3H2} & \ce{H2CO} & \ce{DCO+} \\
		& 6--5 / 3$_{3,0}$--2$_{2,1}$ & 5$_{1,4}$--4$_{2,3}$ / 3$_{3,0}$--2$_{2,1}$ & 3$_{2,2}$--2$_{2,1}$ / 3$_{0,3}$--2$_{0,2}$ & 5--4/3--2 \\
		 & $T_{\rm kin}$ (K) & $T_{\rm kin}$ (K) & $T_{\rm kin}$ (K) & $T_{\rm kin}$ (K) \\
		\hline
		Per18 & 23 $\pm$ 3 & 21 $\pm$ 4 & 28 $\pm$ 3 & 20 $\pm$ 5 \\
		Per21 & 14 $\pm$ 1 & 67 $\pm$ 41 & 29 $\pm$ 2 & 26 $\pm$ 8 \\
		Per49 & 28 $\pm$ 7 & < 71 & 32 $\pm$ 3 & 16 $\pm$ 4 \\
		\hline
		IRAS7 & 18 $\pm$ 2 & 54 $\pm$ 30 & 24 $\pm$ 1 & 17 $\pm$ 3 \\
		\hline
	\end{tabular}
	\label{tab:IRAS7temp}
\end{table*}

\section{Observed molecular line emission}
\label{app:width}
In this appendix the peak antenna temperatures, RMS noise, measured line widths and integrated fluxes are listed for the detected molecular lines from the APEX observations.
These values were obtained with a simple Gaussian fit of the detected emission line, with no fixed parameters, and binning 4 channels to obtain a velocity resolution of 0.4 km~s$^{-1}$. 

\begin{figure*}
	\centering
	\includegraphics[width=0.7\textwidth]{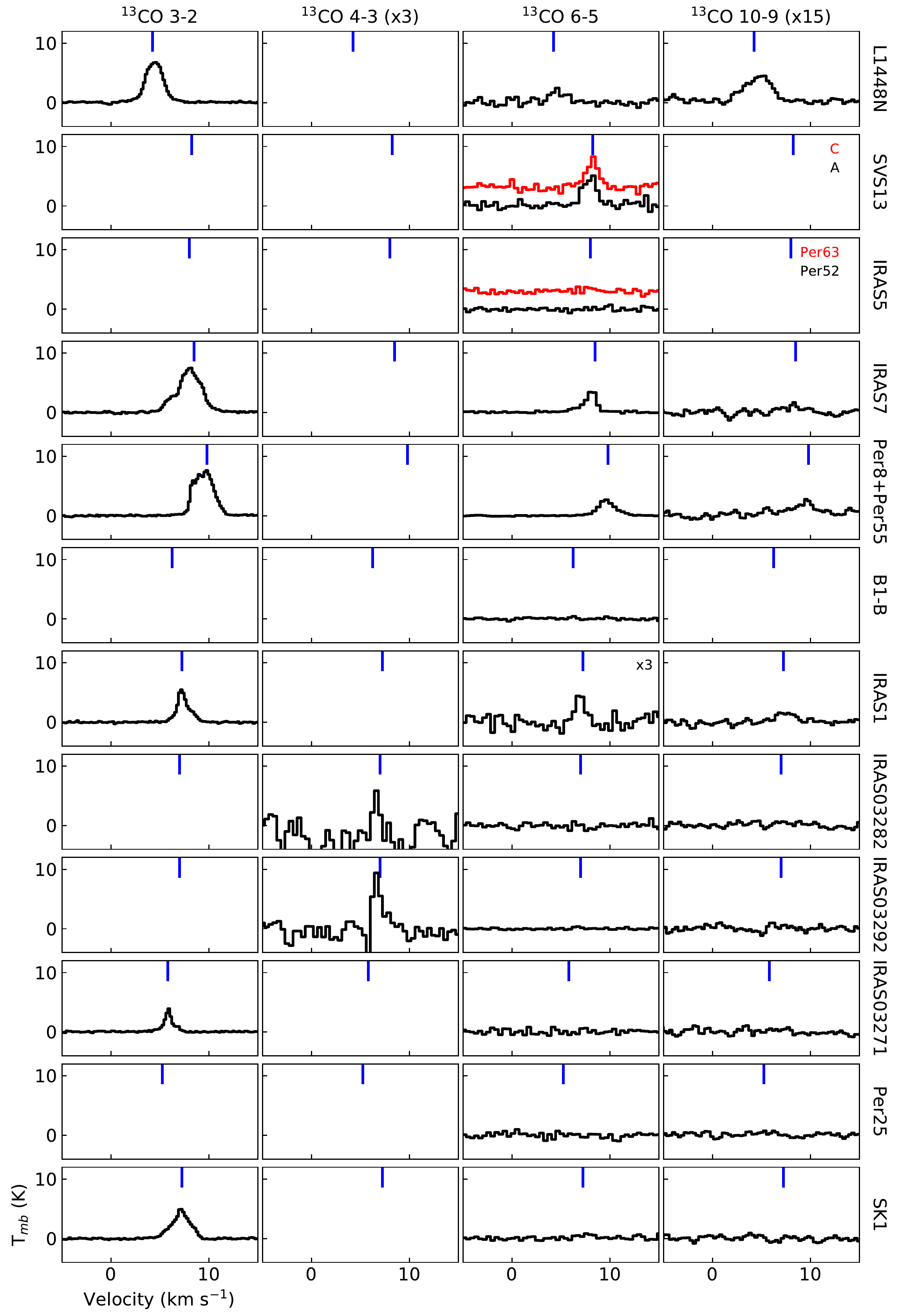}
	\caption{\ce{^{13}CO} spectra observed with JCMT (3--2), APEX (6--5) and \textit{Herschel} HIFI (10--9). The spectra of JCMT and APEX are smoothed to a resolution of 19.25$\arcsec$, the HPBW of the HIFI observations. The short blue lines at the top of each plot indicate the systemic velocity of the system. Note that the \ce{^{13}CO} 4--3 and 10--9 spectra are multiplied by a factor of 3 and 15, respectively. The 6--5 spectra for NGC1333 IRAS1 is also multiplied by a factor of 3.}
	\label{fig:13co}
\end{figure*}

Figure~\ref{fig:13co} shows the spectra of \ce{^{13}CO} in different transitions. The 3--2 transition was observed with the JCMT. Both the 4--3 and 6--5 transitions were observed with APEX using FLASH, CHAMP+ and SEPIA B9. The 10--9 transition was observed with \textit{Herschel} HIFI in the WISH program. All spectra were smoothed to an angular resolution of 19.25$\arcsec$, and the spectra of JCMT and APEX observations were taken with a 19.25$\arcsec$ box. 

\begin{figure*}
	\centering
	\includegraphics[width=0.98\textwidth]{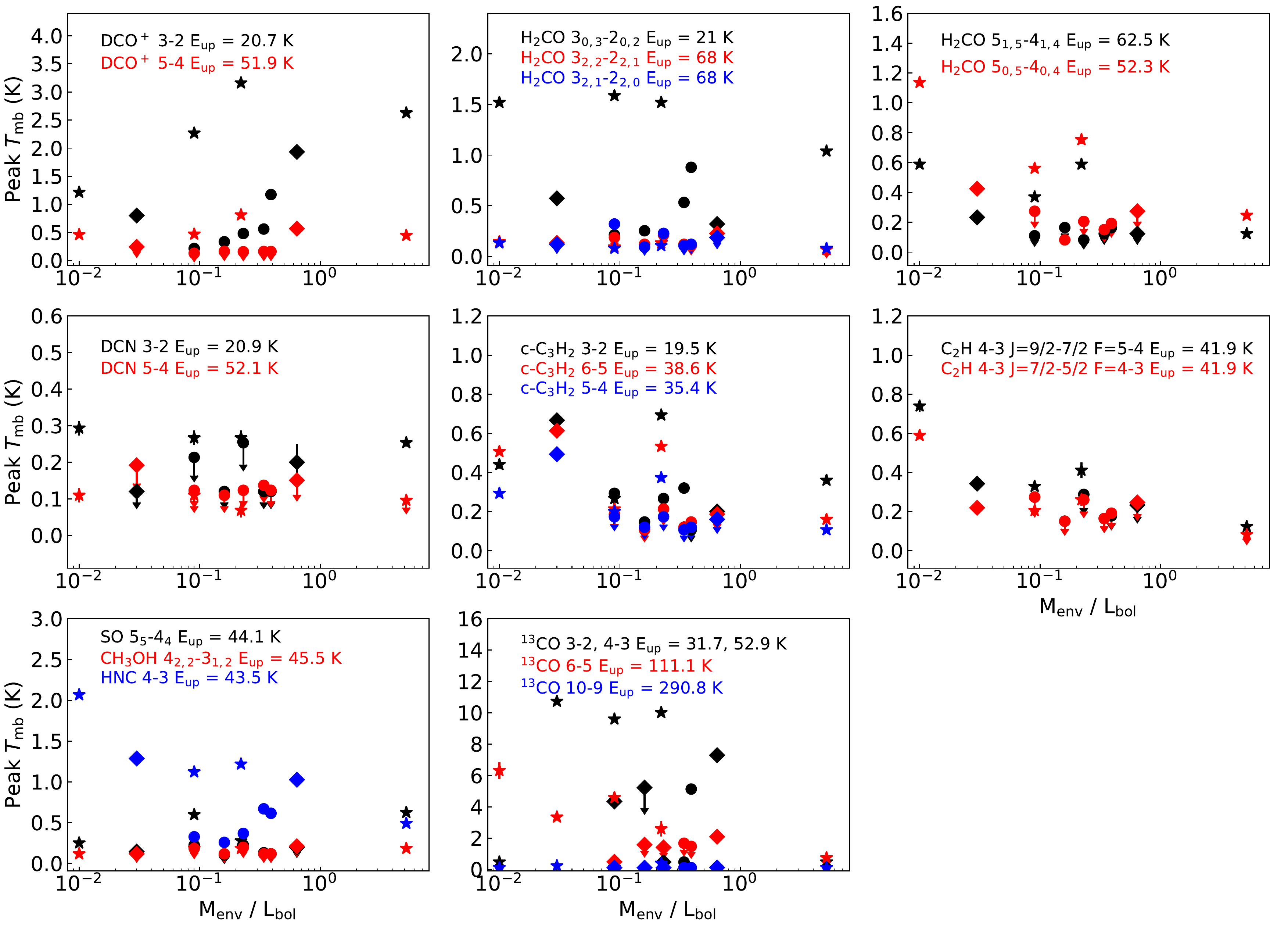}
	\caption{Peak intensities of the observed molecular lines compared to the envelope mass to luminosity ratio of each system. Circles, diamonds and stars show single, close binary and wide multiple protostellar systems, respectively. }
	\label{fig:mlpeak}
\end{figure*}

Figure~\ref{fig:mlpeak} plots the peak antenna temperature from the APEX observations versus the mass to bolometric luminosity ratio $M_{\rm env}/L_{\rm bol}$. No correlation is found between the peak antenna temperatures and $M_{\rm env}/L_{\rm bol}$. This lack of correlation is futher confirmed by the Generalized Kendall's rank correlation (See Sec.~\ref{subsec:stats} and Appendix~\ref{app:stats}), which only finds correlations between Methanol (\ce{CH3OH}) and $M_{\rm env}/L_{\rm bol}$.

\begin{sidewaystable*}
	\centering
	\caption{Peak main beam temperatures for \ce{DCO+} and \ce{DCN}}
	\begin{tabular}{ccccccccccccccccc}
		\hline \hline
	System & \multicolumn{4}{c}{\ce{DCO+} 3–2}       & \multicolumn{4}{c}{\ce{DCO+} 5-4}       & \multicolumn{4}{c}{\ce{DCN} 3–2}       & \multicolumn{4}{c}{\ce{DCN} 5-4}       \\
	& $T_{\rm mb}$ & Noise & Width & Integrated & $T_{\rm mb}$ & Noise & Width & Integrated & $T_{\rm mb}$ & Noise & Width & Integrated & $T_{\rm mb}$ & Noise & Width & Integrated \\
	& mK & mK & km~s$^{-1}$ & mK km~s$^{-1}$ & mK & mK & km~s$^{-1}$ & mK km~s$^{-1}$ & mK & mK & km~s$^{-1}$ & mK km~s$^{-1}$ & mK & mK & km~s$^{-1}$ & mK km~s$^{-1}$ \\
	\hline
	\multicolumn{17}{c}{Wide multiples} \\
	\hline
	L1448N  & 3160 & 31 & 1.13 & 3786.7 & 811 & 33 & 1.04 & 895 & 267 & 25 & 1.63 & 453.3 & 69 & 23 & 1.43 & 105 \\
	SVS13  & 1213 & 23 & 1 & 1306.7 & 461 & 31 & 1.03 & 505 & 293 & 25 & 1.18 & 373.3 & 103 & 28 & 1.82 & 199 \\
	Per63 & 333 & 40 & 0.64 & 226.7 & 162 & 54 & -- & -- & … & 40 & … & … & … & 36 & -- & -- \\
	Per52 & 560 & 37 & 0.66 & 386.7 & 161 & 54 & -- & -- & … & 39 & … & … & … & 47 & -- & -- \\
	IRAS7  & 2267 & 27 & 0.76 & 1866.7 & 468 & 33 & 0.56 & 281 & 267 & 24 & 0.8 & 266.7 & … & 37 & -- & -- \\
	B1-b  & 2627 & 21 & 1.32 & 3680.0 & 446 & 30 & 0.92 & 438 & 253 & 19 & 1.05 & 280.0 & … & 30 & -- & -- \\
	Per8+Per55  & … & … &  &  & … & … & -- & -- & … & … &  &  & … & … & -- & -- \\
	\hline
	\multicolumn{17}{c}{Close binaries} \\
	\hline
	IRAS1  & 800 & 43 & 0.68 & 573.3 & 243 & 81 & -- & -- & … & 40 & … & … & … & 64 & -- & -- \\
	IRAS 03282 & 1933 & 73 & 0.6 & 1200.0 & 567 & 56 & 0.57 & 342 & 200 & 67 & 0.82 & 173.3 & … & 50 & -- & -- \\
	IRAS 03292  & … & … &  &  & … & … & -- & -- & … & … &  &  & … & … & -- & -- \\
	\hline
	\multicolumn{17}{c}{Singles} \\
	\hline
	IRAS 03271 & … & 71 & … & … & … & 43 & -- & -- & … & 69 & … & … & … & 41 & -- & -- \\
	Per25 & 480 & 67 & 0.4 & 200.0 & 156 & 52 & -- & -- & … & 83 & … & … & … & 40 & -- & -- \\
	SK1 & 1173 & 37 & 0.5 & 666.7 & 160 & 53 & -- & -- & … & 40 & … & … & … & 40 & -- & -- \\
		\hline
	\end{tabular}
	\tablefoot{\tablefoottext{a}{In this and subsequent tables, the noise refers to a 0.4 km~s$^{-1}$ velocity bin.}}
	\label{tab:dcodcn}
\end{sidewaystable*}

\begin{sidewaystable*}
	\centering
	\caption{Peak main beam temperatures for \ce{SO}, \ce{CH3OH} and \ce{HNC}}
	\begin{tabular}{ccccccccccccc}
		\hline \hline
		System & \multicolumn{4}{c}{\ce{SO} 5$_{5}$ – 4$_{4}$}       & \multicolumn{4}{c}{\ce{CH3OH} 4$_{2,2}$–3$_{1,2}$}       & \multicolumn{4}{c}{\ce{HNC} 4-3}       \\
		& $T_{\rm mb}$ & Noise & Width & Integrated & $T_{\rm mb}$ & Noise & Width & Integrated & $T_{\rm mb}$ & Noise & Width & Integrated \\
		& mK & mK & km~s$^{-1}$ & K km~s$^{-1}$ & mK & mK & km~s$^{-1}$ & K km~s$^{-1}$ & mK & mK & km~s$^{-1}$ & K km~s$^{-1}$ \\
		\hline
		\multicolumn{13}{c}{Wide multiples} \\
		\hline
		L1448N  & 280 & 27 & 0.99 & 30.7 & 187 & 23 & 1.02 & 200.0 & 1219 & 24 & 2.00 & 2592 \\
		SVS13  & 253 & 23 & 1.5 & 413.3 & 120 & 21 & 1.1 & 133.3 & 2070 & 26 & 1.32 & 2898 \\
		Per63 & … & 35 & … & … & … & 40 & … & … & 256 & 46 & 0.78 & 212 \\
		Per52 & … & 45 & … & … & … & 39 & … & … & 667 & 40 & 0.81 & 573 \\
		IRAS7  & 600 & 23 & 0.74 & 480.0 & 187 & 23 & 0.82 & 160.0 & 1120 & 30 & 0.90 & 1068 \\
		B1-b  & 627 & 21 & 0.85 & 573.3 & 187 & 23 & 0.8 & 160.0 & 490 & 24 & 1.34 & 699 \\
		Per8+Per55  & … & … &  &  & … & … &  &  & -- & -- & … & … \\
		\hline
		\multicolumn{13}{c}{Close binaries} \\
		\hline
		IRAS1  & 147 & 39 & 1.4 & 266.7 & … & 39 & … & … & 1284 & 58 & 0.89 & 1212 \\
		IRAS 03282 & … & 67 & … & … & … & 71 & … & … & 1030 & 44 & 0.84 & 917 \\
		IRAS 03292  & … & … &  &  & … & … &  &  & -- & -- & … & … \\
		\hline
		\multicolumn{13}{c}{Singles} \\
		\hline
		IRAS 03271 & … & 77 & … & … & … & 60 & … & … & 324 & 47 & 1.03 & 357 \\
		Per25 & … & 75 & … & … & … & 65 & … & … & 370 & 38 & 1.07 & 422 \\
		SK1 & … & 39 & … & … & … & 40 & … & … & 611 & 46 & 0.70 & 457 \\
		\hline
	\end{tabular}
	\label{tab:soch3ohhnc}
\end{sidewaystable*}

\begin{sidewaystable*}
	\centering
	\caption{Peak main beam temperatures for \ce{c-C3H2}}
	\begin{tabular}{ccccccccccccc}
		\hline \hline
		System & \multicolumn{4}{c}{\ce{c-C3H2} 3$_{3,0}$--2$_{2,1}$}       & \multicolumn{4}{c}{\ce{c-C3H2} 6–-5}       & \multicolumn{4}{c}{\ce{c-C3H2} 5$_{1,4}$--4$_{2,3}$}       \\
		& Peak & Noise & Width & Integrated & Peak & Noise & Width & Integrated & Peak & Noise & Width & Integrated \\
		& mK & mK & km~s$^{-1}$ & K km~s$^{-1}$ & mK & mK & km~s$^{-1}$ & K km~s$^{-1}$ & mK & mK & km~s$^{-1}$ & K km~s$^{-1}$ \\
		\hline
		\multicolumn{13}{c}{Wide multiples} \\
		\hline
		L1448N  & 693 & 25 & 1.56 & 1146.7 & 533 & 25 & 1.65 & 946.7 & 373 & 23 & 1.68 & 666.7 \\
		SVS13  & 440 & 23 & 0.99 & 453.3 & 507 & 19 & 1.1 & 586.7 & 293 & 23 & 1.3 & 400.0 \\
		IRAS5 Per63 & … & 47 & … & … & … & 35 & … & … & … & 40 & … & … \\
		IRAS5 Per52 & 320 & 40 & 0.5 & 173.3 & 121 & 36 & 0.94 & 121.3 & … & 35 & … & … \\
		IRAS7  & 267 & 24 & 0.7 & 200.0 & 213 & 24 & 0.97 & 226.7 & 200 & 20 & 0.59 & 121.3 \\
		B1-b  & 360 & 23 & 1.33 & 506.7 & 160 & 19 & 0.99 & 173.3 & 107 & 19 & 1.19 & 13.3 \\
		Per8+Per55  & … & … &  &  & … & … &  &  & … & … &  &  \\
		\hline
		\multicolumn{13}{c}{Close binaries} \\
		\hline
		IRAS1  & 667 & 44 & 0.66 & 466.7 & 613 & 40 & 0.75 & 480.0 & 493 & 33 & 0.62 & 333.3 \\
		IRAS 03282  & … & 68 & … & … & … & 61 & … & … & … & 55 & … & … \\
		IRAS 03292  & … & … &  &  & … & … &  &  & … & … &  &  \\
		\hline
		\multicolumn{13}{c}{Singles} \\
		\hline
		IRAS 03271  & … & 97 & … & … & … & 61 & … & … & … & 56 & … & … \\
		Per25  & … & 87 & … & … & … & 69 & … & … & … & 59 & … & … \\
		SK1  & … & 35 & … & … & … & 47 & … & … & … & 40 & … & … \\
		\hline
	\end{tabular}
	\label{tab:c3h2}
\end{sidewaystable*}

\begin{sidewaystable*}
	\centering
	\caption{Peak main beam temperatures for \ce{C2H}}
	\begin{tabular}{ccccccccccccccccc}
		\hline \hline
		System & \multicolumn{4}{c}{\ce{C2H} 4-3 J=9/2 – 7/2 f=5-4}       & \multicolumn{4}{c}{\ce{C2H} 4-3 J=9/2 – 7/2 f=4-3}       & \multicolumn{4}{c}{\ce{C2H} 4-3 J=7/2 – 5/2 f=4-3}       & \multicolumn{4}{c}{\ce{C2H} 4-3 J=7/2 – 5/2 f=3-2}       \\
		& $T_{\rm mb}$ & Noise & Width & Integrated & $T_{\rm mb}$ & Noise & Width & Integrated & $T_{\rm mb}$ & Noise & Width & Integrated & $T_{\rm mb}$ & Noise & Width & Integrated \\
		& mK & mK & km~s$^{-1}$ & mK km~s$^{-1}$ & mK & mK & km~s$^{-1}$ & mK km~s$^{-1}$ & mK & mK & km~s$^{-1}$ & mK km~s$^{-1}$ & mK & mK & km~s$^{-1}$ & mK km~s$^{-1}$ \\
		\hline
		\multicolumn{17}{c}{Wide multiples} \\
		\hline
		L1448N  & 412 & 53 & 2.14 & 937 & 263 & 53 & 2.47 & 693 & 254 & 46 & 2.16 & 584 & 229 & 46 & 2.61 & 637 \\
		SVS13  & 742 & 35 & 1.18 & 928 & 608 & 35 & 1.06 & 688 & 594 & 29 & 1.11 & 703 & 431 & 29 & 1.31 & 602 \\
		Per63 & 0 & 49 & -- & -- & 0 & 49 & -- & -- & 0 & 50 & -- & -- & 0 & 50 & -- & -- \\
		Per52 & 0 & 56 & -- & -- & 0 & 56 & -- & -- & 0 & 53 & -- & -- & 0 & 53 & -- & -- \\
		IRAS7  & 335 & 24 & 0.61 & 216 & 276 & 24 & 1.21 & 356 & 208 & 29 & 0.65 & 143 & 221 & 29 & 0.99 & 232 \\
		B1-b  & 126 & 24 & 0.80 & 108 & 84 & 24 & 0.04 & 840 & 0 & 25 & -- & -- & 0 & 25 & -- & -- \\
		Per8+Per55  & -- & -- & -- & -- & -- & -- & -- & -- & -- & -- & -- & -- & -- & -- & -- & -- \\
		\hline
		\multicolumn{17}{c}{Close binaries} \\
		\hline
		IRAS1  & 337 & 39 & 0.86 & 308 & 258 & 39 & 0.81 & 222 & 215 & 37 & 0.97 & 221 & 138 & 37 & 0.92 & 135 \\
		IRAS 03282 & 0 & 78 & -- & -- & 0 & 78 & -- & -- & 0 & 81 & -- & -- & 0 & 81 & -- & -- \\
		IRAS 03292  & -- & -- & -- & -- & -- & -- & -- & -- & -- & -- & -- & -- & -- & -- & -- & -- \\
		\hline
		\multicolumn{17}{c}{Singles} \\
		\hline
		IRAS 03271 & 0 & 92 & -- & -- & 0 & 92 & -- & -- & 0 & 89 & -- & -- & 0 & 89 & -- & -- \\
		Per25 & 0 & 97 & -- & -- & 0 & 97 & -- & -- & 0 & 88 & -- & -- & 0 & 88 & -- & -- \\
		SK1 & 0 & 60 & -- & -- & 0 & 60 & -- & -- & 0 & 63 & -- & -- & 0 & 63 & -- & -- \\
		\hline
	\end{tabular}
	\label{tab:c2h}
\end{sidewaystable*}

\begin{sidewaystable*}
	\centering
	\caption{Peak main beam temperatures for \ce{H2CO} 3--2}
	\begin{tabular}{ccccccccccccc}
		\hline \hline
		System & \multicolumn{4}{c}{\ce{H2CO} 3$_{0,3}$--2$_{0,2}$}       & \multicolumn{4}{c}{\ce{H2CO} 3$_{2,2}$--2$_{2,1}$}       & \multicolumn{4}{c}{\ce{H2CO} 3$_{2,1}$--2$_{2,0}$}       \\
		& Peak & Noise & Width & Integrated & Peak & Noise & Width & Integrated & Peak & Noise & Width & Integrated \\
		& mK & mK & km~s$^{-1}$ & K km~s$^{-1}$ & mK & mK & km~s$^{-1}$ & K km~s$^{-1}$ & mK & mK & km~s$^{-1}$ & K km~s$^{-1}$ \\
		\hline
		\multicolumn{13}{c}{Wide multiples} \\
		\hline
		L1448N  & 1520 & 23 & 1.68 & 666.7 & 132 & 24 & 2.2 & 306.7 & 109 & 23 & 1.67 & 200.0 \\
		SVS13  & 1520 & 27 & 1.45 & 2333.3 & 147 & 20 & 1.5 & 240.0 & 132 & 21 & 2 & 293.3 \\
		IRAS5 Per63 & 253 & 37 & 1.32 & 360.0 & … & 40 & … & … & … & 32 & … & … \\
		IRAS5 Per52 & 533 & 40 & 0.76 & 400.0 & … & 40 & … & … & … & 37 & … & … \\
		IRAS7  & 1587 & 27 & 1.16 & 1946.7 & 97 & 21 & 1.3 & 133.3 & 80 & 23 & 1.3 & 116.0 \\
		B1-b  & 1040 & 19 & 1.3 & 1453.3 & … & 24 & … & … & 84 & 19 & 0.83 & 73.3 \\
		Per8+Per55  & … & … &  &  & … & … &  &  & … & … &  &  \\
		\hline
		\multicolumn{13}{c}{Close binaries} \\
		\hline
		IRAS1  & 573 & 36 & 0.86 & 520.0 & … & 45 & … & … & … & 41 & … & … \\
		IRAS 03282  & 320 & 60 & 0.97 & 333.3 & … & 73 & … & … & … & 61 & … & … \\
		IRAS 03292  & … & … &  &  & … & … &  &  & … & … &  &  \\
		\hline
		\multicolumn{13}{c}{Singles} \\
		\hline
		IRAS 03271  & … & 69 & … & … & … & 64 & … & … & … & 107 & … & … \\
		Per25  & … & 73 & … & … & … & 69 & … & … & … & 73 & … & … \\
		SK1  & 880 & 43 & 0.95 & 89.3 & … & 35 & … & … & … & 41 & … & … \\
		\hline
	\end{tabular}
	\label{tab:h2co_3-2}
\end{sidewaystable*}

\begin{table*}
	\centering
	\caption{Peak main beam temperatures for \ce{H2CO} 5--4}
	\begin{tabular}{ccccccccc}
		\hline \hline
		System & \multicolumn{4}{c}{5$_{1,5}$-4$_{1,4}$}       & \multicolumn{4}{c}{5$_{0,5}$-4$_{0,4}$}       \\
		& $T_{\rm mb}$ & Noise & Width & Integrated & $T_{\rm mb}$ & Noise & Width & Integrated \\
		& mK & mK & km~s$^{-1}$ & mK km~s$^{-1}$ & mK & mK & km~s$^{-1}$ & mK km~s$^{-1}$ \\
		\hline
		\multicolumn{9}{c}{Wide multiples} \\
		\hline
		L1448N  & 595 & 21 & 2.00 & 1265 & 758 & 40 & 2.03 & 1639 \\
		SVS13  & 594 & 23 & 1.80 & 1136 & 1130 & 27 & 1.38 & 1660 \\
		Per63 & … & 55 & … & … & 82 & 29 & 3.49 & 303 \\
		Per52 & … & 40 & … & … & … & 48 & … & … \\
		IRAS7  & 374 & 33 & 0.98 & 388 & 568 & 32 & 1.22 & 736 \\
		B1-b  & 127 & 19 & 1.92 & 260 & 252 & 25 & 2.52 & 677 \\
		Per8+Per55  & … & … & … & … & … & … & … & … \\
		\hline
		\multicolumn{9}{c}{Close binaries} \\
		\hline
		IRAS1  & 230 & 57 & 2.19 & 535 & 423 & 41 & 1.08 & 489 \\
		IRAS 03282 & … & 42 & 0.00 & 0 & … & 92 & … & … \\
		IRAS 03292  & … & … & … & … & … & … & … & … \\
		\hline
		\multicolumn{9}{c}{Singles} \\
		\hline
		IRAS 03271 & … & 36 & … & … & … & 90 & … & … \\
		Per25 & … & 28 & … & … & … & 69 & … & … \\
		SK1 & 161 & 40 & 0.92 & 158 & … & 62 & … & … \\
		\hline
	\end{tabular}
	\label{tab:h2co_5-4}
\end{table*}

\begin{sidewaystable*}
	\centering
	\caption{Peak main beam temperatures for \ce{^{13}CO}}
	\begin{tabular}{ccccccccccccccccc}
		\hline \hline
		System & \multicolumn{4}{c}{3--2\tablefootmark{a}} & \multicolumn{4}{c}{4--3\tablefootmark{b}} & \multicolumn{4}{c}{6--5\tablefootmark{b}} & \multicolumn{4}{c}{10--9\tablefootmark{c}} \\
		& $T_{\rm mb}$ & Noise & Width & Integrated & $T_{\rm mb}$ & Noise & Width & Integrated & $T_{\rm mb}$ & Noise & Width & Integrated & $T_{\rm mb}$ & Noise & Width & Integrated \\
		& K & K & km~s$^{-1}$ & K km~s$^{-1}$ & K & K & km~s$^{-1}$ & K km~s$^{-1}$ & K & K & km~s$^{-1}$ & K km~s$^{-1}$ & K & K & km~s$^{-1}$ & K km~s$^{-1}$ \\
		\hline
		\multicolumn{17}{c}{Wide multiples} \\
		\hline
		L1448N & 7.51 & 0.12 & 2.15 & 17.22 &  &  &  &  & 1.95 & 0.49 & 2.03 & 4.22 & 0.30 & 0.03 &  & 1.0034 \\
		SVS13-A &  &  &  &  &  &  &  &  & 4.83 & 0.56 & 1.71 & 8.78 &  &  &  &  \\
		SVS13-C &  &  &  &  &  &  &  &  & 4.66 & 0.50 & 1.57 & 7.78 &  &  &  &  \\
		IRAS7\tablefootmark{d}& 7.20 & 0.13 & 2.93 & 22.46 &  &  &  &  & 3.44 & 0.14 & 1.41 & 5.16 & 0.11 & 0.03 &  & 0.17 \\
		B1-B &  &  &  &  &  &  &  &  &...& 0.19 &...&...&  &  &  &  \\
		PER8+PER55 & 8.05 & 0.12 & 2.48 & 21.23 &  &  &  &  & 2.51 & 0.06 & 2.07 & 5.54 & 0.18 & 0.03 &  & 0.6882 \\
		\hline
		\multicolumn{17}{c}{Close binaries} \\
		\hline
		IRAS1 & 5.47 & 0.13 & 1.40 & 8.16 &  &  &  &  & 1.57 & 0.33 & 1.10 & 1.84 & 0.11 & 0.03 &  & 0.2244 \\
		IRAS 03282 &  &  &  &  & 1.99 & 0.75 & 0.62 & 1.32 & ... & 0.40 &...&...& ... & 0.03 &...&...\\
		IRAS 03292 &  &  &  &  & 3.26 & 0.44 & 0.84 & 2.91 & ... & 0.12 &...&...& ... & 0.03 &...&...\\
		\hline
		\multicolumn{17}{c}{Singles} \\
		\hline
		IRAS 03271 & 3.85 & 0.13 & 1.06 & 4.35 &  &  &  &  & ... & 0.37 &...&...& 0.09 & 0.03 &...&...\\
		Per25 &  &  &  &  &  &  &  &  & ... & 0.42 &...&...& ... & 0.03 &...&...\\
		SK1 & 4.47 & 0.13 & 2.26 & 10.73 &  &  &  &  & ... & 0.29 &...&...& ... & 0.03 &...&...\\
		\hline
	\end{tabular}
	\tablefoot{Blank spaces indicate that there are no observations for the respective system and transition; three dots (...) indicate non-detection.
		\tablefoottext{a}{Peak antenna temperatures from JCMT observations, smoothed to a beam of 19.25$\arcsec$}
		\tablefoottext{b}{Peak antenna temperatures and noise from APEX observations, averaged over a box of 19.25$\arcsec$.}
		\tablefoottext{c}{Peak antenna temperatures from $Herschel$ HIFI observations with a beam of 19.25$\arcsec$ \citep{mottram2017}.}
		\tablefoottext{d}{IRAS 7 here includes only Per18 and Per21}}
	\label{tab:13co}
\end{sidewaystable*}

\section{Statistical analysis}
\label{app:stats}
The relation between the results of the observations and the system parameters is determined by the Generalized Kendall's rank \citep{isobe1986}. The table in this appendix lists the significance level $p$ and standard normal score $z$ of the rank correlation test. The numbers in bold indicate real correlations ($p~<$ 0.05).

\begin{table*}
	\centering
	\caption{Results of the Generalized Kendall Correlation}
	\begin{tabular}{c c c c c c c c c c}
		\hline \hline
		Molecule & Transition & \multicolumn{2}{c}{Mass}   && \multicolumn{2}{c}{$L_{\rm bol}$}  && \multicolumn{2}{c}{Mass / $L_{\rm bol}$}   \\
		\cline{3-4}
		\cline{6-7}
		\cline{9-10}
		&  & z & p && z & p && z & p \\
		\hline															
		\multicolumn{10}{c}{Peak antenna temperatures} \\														
		\hline															
		\ce{SO}	&	 5$_{5}$--4$_{4}$ 	&	2.34	&	\textbf{0.01}	&&	0.94	&	0.17	&&	-0.23	&	0.41	\\
		\ce{DCO+}	&	 3--2 	&	3.04	&	\textbf{0.00}	&&	0.54	&	0.29	&&	1.09	&	0.14	\\
		\ce{c-C3H2}	&	 3$_{3,0}$--2$_{2,1}$ 	&	0.94	&	0.17	&&	1.56	&	0.06	&&	-1.02	&	0.15	\\
		\ce{DCN}	&	 3--2 	&	1.85	&	\textbf{0.03}	&&	1.85	&	\textbf{0.03}	&&	-0.81	&	0.21	\\
		\ce{c-C3H2}	&	 6--5 	&	1.49	&	0.07	&&	2.58	&	\textbf{0.00}	&&	-1.57	&	0.06	\\
		\ce{c-C3H2}	&	 5$_{1,4}$--4$_{2,3}$ 	&	1.1	&	0.14	&&	3.3	&	\textbf{0.00}	&&	-2.44	&	\textbf{0.01}	\\
		\ce{H2CO}	&	 3$_{0,3}$--2$_{0,2}$ 	&	2.03	&	\textbf{0.02}	&&	0.94	&	0.17	&&	-0.23	&	0.41	\\
		\ce{CH3OH}	&	 4$_{2,2}$--3$_{1,2}$ 	&	0.93	&	0.18	&&	-0.08	&	0.47	&&	1.27	&	0.10	\\
		\ce{H2CO}	&	 3$_{2,2}$--2$_{2,1}$ 	&	-0.55	&	0.29	&&	1.02	&	0.15	&&	-0.63	&	0.26	\\
		\ce{H2CO}	&	 3$_{2,1}$--2$_{2,0}$ 	&	-1.1	&	0.14	&&	0.63	&	0.26	&&	-0.39	&	0.35	\\
		\ce{C2H}	&	 4--3 J=9/2--7/2 F=5--4 	&	1.17	&	0.12	&&	3.35	&	\textbf{0.00}	&&	-2.34	&	\textbf{0.01}	\\
		\ce{C2H}	&	 4--3 J=9/2--7/2 F=4--3 	&	0.23	&	0.41	&&	2.11	&	\textbf{0.02}	&&	-1.96	&	\textbf{0.02}	\\
		\ce{C2H}	&	 4--3 J=7/2--5/2 F=4--3 	&	0.47	&	0.32	&&	2.34	&	\textbf{0.01}	&&	-1.8	&	\textbf{0.04}	\\
		\ce{C2H}	&	 4--3 J=7/2--5/2 F=3--2 	&	-0.08	&	0.47	&&	1.48	&	0.07	&&	-0.94	&	0.17	\\
		\ce{H2CO}	&	 5$_{1,5}$--4$_{1,4}$ 	&	1.74	&	\textbf{0.04}	&&	2.22	&	\textbf{0.01}	&&	-1.51	&	0.07	\\
		\ce{DCO+}	&	 5--4 	&	2.76	&	\textbf{0.00}	&&	1.34	&	0.09	&&	0.16	&	0.44	\\
		\ce{DCN}	&	 5--4 	&	-1.67	&	0.05	&&	-0.56	&	0.29	&&	0.16	&	0.44	\\
		\ce{HNC}	&	 4--3 	&	1.63	&	0.05	&&	1.95	&	\textbf{0.03}	&&	-1.25	&	0.11	\\
		\ce{H2CO}	&	 5$_{0,5}$--4$_{0,4}$ 	&	2.03	&	\textbf{0.02}	&&	3.12	&	\textbf{0.00}	&&	-1.49	&	0.07	\\
		\ce{^{13}CO}	&	3–2 \& 4–3	&	0.08	&	0.47	&&	1.76	&	\textbf{0.04}	&&	-0.64	&	0.26	\\
		\ce{^{13}CO}	&	6–5	&	0.39	&	0.35	&&	2.81	&	\textbf{0.00}	&&	-2.5	&	\textbf{0.01}	\\
		\ce{^{13}CO}	&	10–9	&	0.55	&	0.29	&&	1.92	&	\textbf{0.03}	&&	-0.64	&	0.26	\\
		\hline															
		\multicolumn{10}{c}{Ratios}	\\													
		\hline															
		\ce{DCN}/\ce{DCO+}	&	 3--2 	&	-1.82	&	\textbf{0.03}	&&	-0.18	&	0.43	&&	-1.27	&	0.10	\\
		\ce{DCO+}	&	5-4/3-2	&	-0.9	&	0.18	&&	0.9	&	0.18	&&	-1.44	&	0.08	\\
		\ce{c-C3H2}	&	6--5 / 3$_{3,0}$--2$_{2,1}$	&	-0.19	&	0.43	&&	2.07	&	\textbf{0.02}	&&	-2.44	&	\textbf{0.01}	\\
		\ce{c-C3H2}	&	5$_{1,4}$--4$_{2,3}$ / 3$_{3,0}$--2$_{2,1}$	&	-0.94	&	0.17	&&	0.56	&	0.29	&&	-1.69	&	0.05	\\
		\ce{H2CO}	&	3$_{2,2}$--2$_{2,1}$ / 3$_{0,3}$--2$_{0,2}$	&	-1.46	&	0.07	&&	-0.21	&	0.42	&&	0.21	&	0.42	\\
		\ce{^{13}CO}	&	6–5/3--2	&	0.51	&	0.30	&&	0.94	&	0.17	&&	-0.94	&	0.17	\\
		\ce{^{13}CO}	&	10–9/3--2	&	2.03	&	\textbf{0.02}	&&	0.23	&	0.41	&&	1.6	&	0.05	\\
		\ce{^{13}CO}	&	10–9/6--5	&	0.36	&	0.36	&&	1.53	&	0.06	&&	1.08	&	0.14	\\
		\hline															
	\end{tabular}
	\label{tab:stats}
\end{table*}

\end{appendix}

\end{document}